\documentclass{iopjournal}
\usepackage{lmodern}
\usepackage[backend=biber,style=phys,biblabel=brackets,maxbibnames=4]{biblatex}
\usepackage{subcaption}
\usepackage{colortbl}
\usepackage{lineno}
\usepackage{multirow}
\usepackage{graphicx}
\usepackage{hyperref} 
\usepackage{amssymb}
\usepackage{acronym}
\usepackage{ulem}
\usepackage{comment}
\usepackage{ragged2e}

\usepackage{tikz}
\usepackage{calc}
\def\checkmark{\tikz\fill[scale=0.4](0,.35) -- (.25,0) -- (1,.7) -- (.25,.15) -- cycle;}

\acrodef{GW}[GW]{gravitational-wave}
\acrodef{BBH}[BBH]{binary black hole}
\acrodef{LVK}[LVK]{LIGO-Virgo-KAGRA Collaboration}
\acrodef{O3}[O3]{third observing run}
\acrodef{O4}[O4]{fourth observing run}
\acrodef{O4a}[O4a]{the first part of O4}
\acrodef{CNN}[CNN]{convolutional neural network}
\acrodef{SNR}[SNR]{signal-to-noise ratio}
\acrodef{DQR}[DQR]{Data Quality Report}
\acrodef{LLO}[LLO]{LIGO Livingston Observatory}
\acrodef{LHO}[LHO]{LIGO Hanford Observatory}
\acrodef{DQ}[DQ]{data quality}
\acrodef{LM}[LM]{Low-mass}
\acrodef{HM}[HM]{High-mass}
\acrodef{EHM}[EHM]{Extremely high-mass}
\acrodef{ROC}[ROC]{Receiver Operating Characteristic}
\acrodef{Grad-CAM}[\texttt{Grad-CAM}]{Gradient-weighted Class Activation Mapping}
\acrodef{TAR}[TAR]{True Alarm Rate}
\acrodef{FAR}[FAR]{False Alarm Rate}
\acrodef{AUC}[AUC]{area under the curve}

\begin{document}
\articletype{Paper}

\title{GSpyNetTree-O4: an event validation tool used in the fourth LIGO-Virgo-KAGRA observing run}

\author{Sof\'ia \'Alvarez-L\'opez$^{1,2,*}$\orcid{0009-0003-8040-4936},  Man Leong Chan$^{3}$, Franz S. Herbst$^{3,4}$, Dhatri Raghunathan$^{3}$, Airene Ahuja$^{3}$, Annudesh Liyanage$^{3}$, Julian Ding$^{3}$, Alejandro Garcia-Varela$^{5}$, Raymond Ng$^{6}$, and Jess McIver$^{3}$}

\affil{$^1$ LIGO Laboratory, Massachusetts Institute of Technology, Cambridge, MA 02139, USA}

\affil{$^2$ Kavli Institute for Astrophysics and Space Research and Department of Physics, Massachusetts Institute of Technology, Cambridge, MA 02139, USA}

\affil{$^3$ Department of Physics and Astronomy, University of British Columbia, Vancouver, British Columbia, V6T1Z4, Canada}

\affil{$^4$ Department of Physics, University of Konstanz, 78457, Konstanz, Germany}

\affil{$^5$ Departamento de F\'isica, Universidad de los Andes, 111711, Bogot\'a, Colombia}

\affil{$^6$ Department of Computer Science, University of British Columbia, Vancouver, British Columbia, V6T1Z4, Canada}

\affil{$^*$Author to whom any correspondence should be addressed.}

\email{sofiaal@mit.edu}


\begin{abstract}
The frequent presence of non-Gaussian transient noise, or glitches, in gravitational-wave detector data can affect gravitational-wave searches, parameter estimation, and downstream analyses. To identify and mitigate transient noise near gravitational-wave candidates in a timely manner, the LIGO-Virgo-KAGRA Collaboration employs the Data Quality Report. In the fourth observing run, \texttt{GSpyNetTree-O4} was deployed within this framework as a tool for glitch classification and event validation. We describe \texttt{GSpyNetTree-O4} and the main developments relative to its predecessor, \texttt{GSpyNetTree}. The most important update was a new architecture that allowed the simultaneous identification of glitches and gravitational-wave signals when both were present in the same input. We also expanded and augmented the training set with examples in which simulated gravitational-wave signals overlapped with real glitches, and applied 60\,Hz calibration corrections to better match the data expected during the fourth observing run. On test data, the low-mass, high-mass, and extremely high-mass classifiers identified 97.9\%, 97.7\%, and 95.4\% of glitches, respectively. Among samples without a glitch, including gravitational-wave-only and No Glitch samples, the classifiers correctly reported no data-quality issues in 97.1\%, 96.6\%, and 96.0\% of cases, respectively. We further assessed the robustness of \texttt{GSpyNetTree-O4} on unseen glitch morphologies, a small set of Virgo glitches from the fourth observing run, and different choices of the $Q$-value used to construct the time-frequency inputs. \texttt{GSpyNetTree-O4} was successfully deployed as a Data Quality Report tool and increased automation in gravitational-wave event validation workflows.
\end{abstract}

\section{Introduction}
Recent upgrades in instrumentation have steadily improved the sensitivity of ground-based \ac{GW} detectors, including LIGO~\cite{aLIGO, PhysRevD.111.062002}, Virgo~\cite{AdVirgo, Acernese_2023}, and KAGRA~\cite{10.1093/ptep/ptaa125, 10.1093/ptep/ptac093}. Up to the second part of the fourth observing run, the \ac{LVK} has reported more than 250 \ac{GW} signals~\cite{o2,o3a,o3b,gwtc-4}, with additional candidates reported by external teams~\cite{2OGC, 3OGC, PhysRevD.104.063030, PhysRevD.101.083030}. With analyses of the remaining \ac{O4} data still ongoing\footnote{\url{https://gracedb.ligo.org/superevents/public/O4/}}, the catalog is expected to expand further. Future observing runs with increased detector sensitivity may also enable observations of \ac{GW} sources that have so far remained undetected, such as core-collapse supernovae and spinning neutron stars.

However, \ac{GW} detection is still challenging. The extreme sensitivity of ground-based interferometers makes these detectors prone to a variety of transient noise features, known as \textit{glitches}, which can limit astrophysical analyses~\cite{1stnoise, Abbott_2020kk}. These non-Gaussian, short-duration artifacts can arise from environmental and instrumental disturbances and often occur frequently~\cite{derek}. For example, during \ac{O4a}, a rate of approximately one glitch per minute\footnote{Including glitches with a signal-to-noise ratio above 6.5.} was observed in the LIGO detectors~\cite{LIGO:2024kkz}. Such a high glitch rate increases the likelihood that \ac{GW} candidates occur in close proximity to glitches. Indeed, this was the case for more than 30\% of the \ac{GW} candidates reported in the recent GWTC-4 catalog~\cite{gwtc-4}, as well as for the binary neutron star merger GW170817~\cite{gw170817}. Furthermore, it has been demonstrated across a broad range of compact-binary masses that glitches occurring close in time to \ac{GW} signals can distort source property estimation~\cite{massinger, Canton:2013joa, PhysRevD.110.122002, PhysRevD.106.042006, macas, PhysRevD.98.084016, curious_case, powell_jade, Udall:2024ovp}. 

To address these challenges, the \ac{LVK} performs event validation to identify and mitigate glitches that could impact downstream analyses~\cite{1stnoise, derek, LIGO:2024kkz}. To manage the increasing rate of candidate signals, the \ac{LVK} implemented the \ac{DQR}~\cite{LIGOScientific:2026ifv, Davis:2026kkz} to automate the production of event validation results. The DQR compiles results from a variety of tools designed to assess the quality of detector data in the vicinity of a \ac{GW} candidate. Despite the success of this approach in reducing validation time in previous observing runs~\cite{derek}, pre-existing tools had limitations that hindered further automation. For example, \texttt{GlitchFind}~\cite{Vazsonyi:2022jul} can identify cases where glitches elevate detector noise relative to surrounding data stretches, but is less effective when the glitch rate is elevated for a longer period, e.g., during light scattering, which can last for hours when local ground motion is elevated~\cite{sidScattering}. Similarly, \texttt{iDQ}~\cite{PhysRevD.88.062003, Essick_2013, idq} is extremely effective in cases where one or more sensors witness a glitch, but is otherwise blind to glitches without witnesses, as is the case for most recurring short-duration glitches, e.g., blips~\cite{derek, LIGO:2024kkz}.

\textcite{gspynettree1} previously introduced \texttt{GSpyNetTree}, a deep learning algorithm developed from the glitch classification framework \texttt{Gravity Spy}~\cite{gspy,Zevin:2023rmt}, and demonstrated its potential to effectively mitigate these challenges and increase automation in the event validation process~\cite{gspynettree1}. \texttt{GSpyNetTree} consisted of an ensemble of \acp{CNN} that classified inputs as one of several glitch classes, a \ac{GW} signal, or No Glitch. Its performance, however, decreased when the background-noise properties differed from those represented in the training set, reflecting the limited variation in background noise within the training data. It also performed less reliably on glitch morphologies that were absent or poorly represented in the training set, including Low-frequency Lines, which became a prominent glitch class in \ac{O4}~\cite{LIGO:2024kkz}. Other development pathways for \texttt{GSpyNetTree} have also been explored, including \texttt{GSpyNetTree-S}~\cite{Chan:2025zab}, a segmentation algorithm built on the \texttt{You Only Look Once}~\cite{yolo} architectures, and \texttt{GSpyNetTree-Bayes}~\cite{gsnt_bayes}, which incorporates Bayesian convolutional neural networks for uncertainty estimation.

In this paper, we describe \texttt{GSpyNetTree-O4}, which was deployed by the \ac{LVK} at the beginning of \ac{O4} for glitch classification and event validation and operated throughout the run. It was implemented as part of the \texttt{dqr-tasks} package\footnote{Publicly available at \url{https://pypi.org/project/dqrtasks/}} within the \texttt{DQRbuild} toolkit~\cite{Davis:2026kkz}. We also present the principal improvements relative to the original algorithm. In Section~\ref{sec:methods}, we describe the methodology used to improve \texttt{GSpyNetTree-O4}'s performance relative to the original \texttt{GSpyNetTree}. In Section~\ref{sec:results}, we present performance results for the version of \texttt{GSpyNetTree-O4} that ran in production in the \ac{LVK}'s \ac{DQR} and enabled validation of events in the LVK's most recently published catalogs~\cite{gwtc-4, LIGOScientific:2026wfs}, as well as future LVK catalogs expected within the year. In Section~\ref{sec:furtherstudies}, we describe pathways for future improvement toward full automation of \ac{GW} candidate validation. In Section~\ref{sec:conclusions}, we summarize our conclusions.

\section{\texttt{GSpyNetTree-O4} improvements}\label{sec:methods}

\texttt{GSpyNetTree-O4} follows the architectural philosophy of its predecessor~\cite{gspynettree1}. It consists of a decision tree of three InceptionV3~\cite{inceptionv3} \acp{CNN}, trained on simulated GW signals and morphologically similar glitches. Both \texttt{GSpyNetTree} and \texttt{GSpyNetTree-O4} use a feature set similar to \texttt{Gravity Spy}~\cite{gspy, jane, gspyO3, Zevin:2023rmt, Wu_2025}: each input sample is composed of a $2\times2$ matrix of spectrogram visualizations that are 0.5, 1, 2, and 4 seconds in duration. To capture the different morphologies between short-duration and long-duration glitches and \ac{GW} sources, as recommended by~\textcite{jarov}, different models were deployed depending on a \ac{GW} candidate's estimated total mass, $M$: the \ac{LM} classifier handled events with $M < 50\; M_\odot$, the \ac{HM} classifier covered events in the $50\; M_\odot \leq M < 250\; M_\odot$ range, and the \ac{EHM} classifier was used for events with $M \geq 250\; M_\odot$. 

\begin{figure}
\centering
\includegraphics[width=\linewidth]{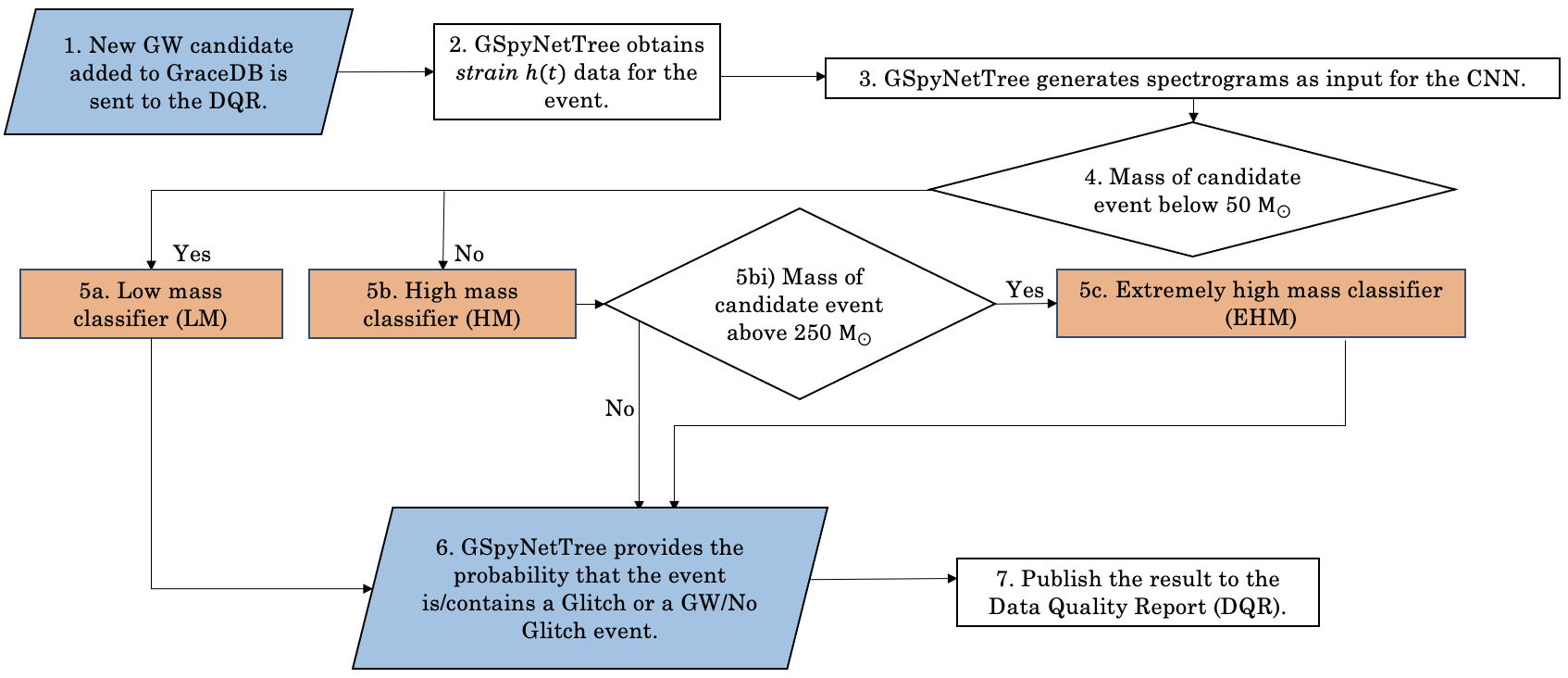}
\caption{\texttt{GSpyNetTree-O4}'s architecture (cf. Figure 3 in \textcite{gspynettree1}). The addition of a new candidate to GraceDB~\cite{gracedb} initiates the \ac{DQR} workflow~\cite{Davis:2026kkz}, which runs a suite of event-validation tasks, including \texttt{GSpyNetTree-O4}. \texttt{GSpyNetTree-O4} obtains strain $h(t)$ data associated with the event and generates a $2\times2$ grid of spectrograms with durations of $0.5, 1, 2,$ and $4$ seconds. These time-frequency visualizations are sent to one of three classifiers based on the candidate's estimated total mass. Each classifier outputs probabilities indicating whether the event is or contains one of the target glitch classes, a \ac{GW}, or no glitch.  Finally, the result is published in the form of a webpage to the \ac{DQR}, accessible to \ac{LVK} event validators.}
\label{fig:arch}
\end{figure}

For \ac{O4}, \texttt{GSpyNetTree-O4} was deployed on LIGO data as part of the \ac{LVK} event-validation process~\cite{LIGO:2024kkz}. When a new \ac{GW} candidate was uploaded to GraceDB\footnote{\url{https://gracedb.ligo.org}}~\cite{gracedb}, the \ac{DQR} triggered \texttt{GSpyNetTree-O4}'s analysis~\cite{Davis:2026kkz}. \texttt{GSpyNetTree-O4} applied the appropriate model (\ac{LM}, \ac{HM}, or \ac{EHM}) depending on the total mass of the candidate, which was estimated using the low-latency pipelines PyCBC Live~\cite{pycbc1, pycbc2}, MBTA~\cite{mbta1, mbta2}, GstLAL~\cite{gstlal1, gstlal2, gstlal3, gstlal4}, or SPIIR~\cite{SPIIR1, SPIIR2} and reported to GraceDB. The selected model then predicted the probability of the presence of a glitch in detector data surrounding the \ac{GW} candidate. The results were published on a \ac{DQR} webpage, following the architecture shown in Figure \ref{fig:arch}.  \texttt{GSpyNetTree-O4}'s output was used by \ac{LVK} event validators to assess the quality of candidate events and inform the need for glitch mitigation, along with the output of other \ac{DQR} tasks like \texttt{GlitchFind} and \texttt{Omega Overlap}~\cite{LIGO:2024kkz, Davis:2026kkz}.

Building on \texttt{GSpyNetTree}'s original architecture, we developed \texttt{GSpyNetTree-O4} by implementing several key recommendations from \textcite{gspynettree1}. These enhancements made \texttt{GSpyNetTree-O4} more robust to variations in background noise and \acp{GW} occurring in close proximity to glitches. In the following sections, we expand on the implementation of these improvements, along with other upgrades introduced in \texttt{GSpyNetTree-O4}.

\subsection{Construction of a representative and comprehensive glitch sample set }\label{sec:o4prep}
\begin{figure}
\centering
\includegraphics[width=\linewidth]{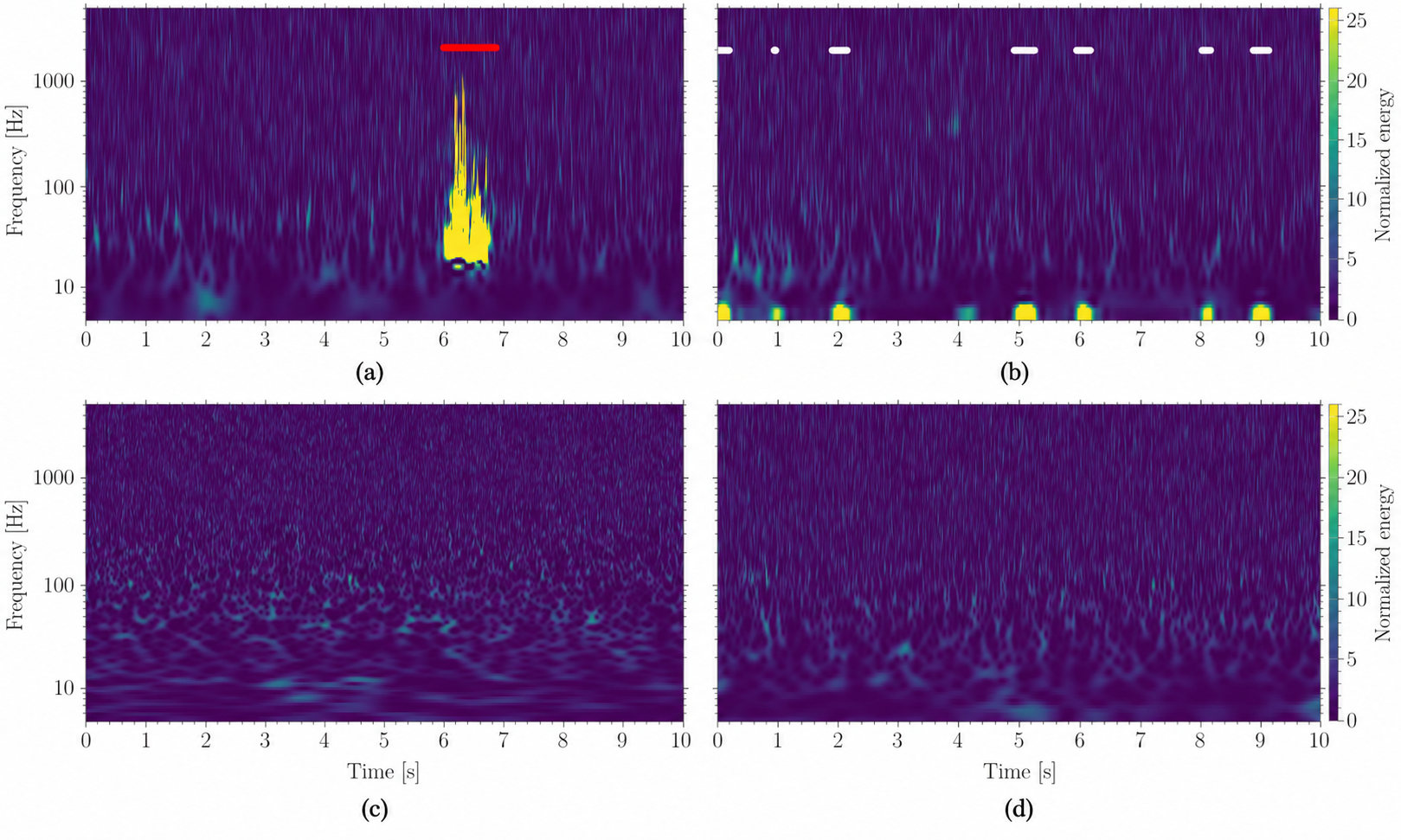}
\caption{\textit{Top panel:} Spectrograms of noisy detector data: (a) shows a Koi Fish glitch identified at \ac{LLO} in O3. The time-frequency tiles flagged by the algorithm described in Section~\ref{sec:noiseRobustness} are shown in red. This glitch was subsequently included in \texttt{GSpyNetTree-O4}'s training set. (b) shows a Virgo spectrogram with abundant low-frequency glitches below $10\;\mathrm{Hz}$. The time-frequency tiles flagged by the algorithm are shown in white. \textit{Bottom panel:} Spectrograms of clean detector background times from (c) \ac{LHO} and (d) \ac{LLO} during \ac{O3}. In \texttt{GSpyNetTree-O4}, these clean times were used to inject simulated \ac{GW} signals.
}
\label{fig:specs}
\end{figure}

In \textcite{gspynettree1}, all glitches in the training set were obtained through LIGO-DV web~\cite{ligodv}, and the Gravitational Wave Open Science Center (GWOSC)~\cite{gwosc1, gwosc2, gwosc3}. In this work, we continued to use these glitches for \texttt{GSpyNetTree-O4}, but we increased the number of glitch samples in the training set in two ways. First, we included glitches found in Virgo data from the \ac{O3}. Although \texttt{GSpyNetTree-O4} was deployed on LIGO data, Virgo glitches in the classes considered here exhibit morphologies similar to their LIGO counterparts. Including these samples allowed us to improve robustness to subtle morphological variations within the same glitch category. Second, we increased the number of glitches in our training set by including 193 manually verified glitches from non-quiet data segments flagged by the algorithm described in Section~\ref{sec:noiseRobustness}. We included only glitches with excess power above $10\,\mathrm{Hz}$. Figure~\ref{fig:specs}a shows an example of a Koi Fish glitch identified in this way during \ac{O3} at \ac{LLO}. We also found a considerable number of Virgo noise transients below $10\,\mathrm{Hz}$, one of which is shown in Figure~\ref{fig:specs}b. We did not include these low-frequency transients because \texttt{GSpyNetTree-O4} was developed primarily to run on LIGO data. A future version designed for Virgo data will require an expanded training set that includes these types of transients.

\begin{table}
\centering
\small
\begin{tabular}{lcccc}
\hline
\textbf{GSpyNetTree label} &
  \textbf{\begin{tabular}[c]{@{}c@{}}Total number \\ of samples\end{tabular}} &
  \textbf{\begin{tabular}[c]{@{}c@{}}Low-mass \\ classifier\end{tabular}} &
  \textbf{\begin{tabular}[c]{@{}c@{}}High-mass \\ classifier\end{tabular}} &
  \textbf{\begin{tabular}[c]{@{}c@{}}Extremely \\ high-mass \\ classifier\end{tabular}} \\
\hline
Blip & 7725 & \checkmark & \checkmark & \checkmark \\ 
Low-frequency Blip & 5000 & \checkmark & \checkmark & \checkmark \\ 
Fast Scattering & 5305 & \checkmark & \checkmark & \checkmark \\
Light Scattering & 4215 & \checkmark & \checkmark & \checkmark \\ 
Low-frequency Lines & 5250 & \checkmark & \checkmark & \checkmark \\ 
No Glitch & 14610 & \checkmark & \checkmark & \checkmark \\ 
Koi Fish & 5330 & \checkmark & \checkmark & \\ 
Scratchy & 5305 & \checkmark & & \\ 
Tomte & 7085 & & \checkmark & \\ 
GW ($3-50\;M_\odot$) & 14218 & 14218 & & \\ 
GW ($50-250\;M_\odot$) & 17489 & & 17489 & \\ 
GW ($250-350\;M_\odot$) & 12493 & & & 12493 \\ 
\hline
\multicolumn{5}{c}{\textbf{Overlapping samples}} \\
\hline
Blip + \ac{GW} & 36209 & 12070 & 12069 & 12070 \\ 
Low-frequency Blip + \ac{GW} & 23880 & 7960 & 7960 & 7960 \\ 
Fast Scattering + \ac{GW} & 37050 & 12350 & 12350 & 12350 \\ 
Light Scattering + \ac{GW} & 35788 & 11928 & 11930 & 11930 \\ 
Koi Fish + \ac{GW} & 9600 & 4800 & 4800 & 0 \\ 
Scratchy + \ac{GW} & 5305 & 5305 & 0 & 0 \\ 
Tomte + \ac{GW} & 11045 & 0 & 11045 & 0 \\ 
\hline
\end{tabular}
\caption{Number of samples per class for the \ac{LM}, \ac{HM}, and \ac{EHM} classifiers. These total numbers include the following data augmentation techniques: 1) doubling the number of LIGO samples by applying the nonlinear $60\;\mathrm{Hz}$ AC power artifact subtraction for noise robustness (described in Section~\ref{sec:overlapping}), and 2) applying time-offset augmentation, which adds four more examples by shifting the glitch or \ac{GW} from $t=0\,\mathrm{s}$.}
\label{tab:class size}
\end{table}

\begin{figure}
\centering
\captionsetup[subfigure]{font=small}
     \begin{subfigure}[b]{\textwidth}
         \includegraphics[width=\textwidth]{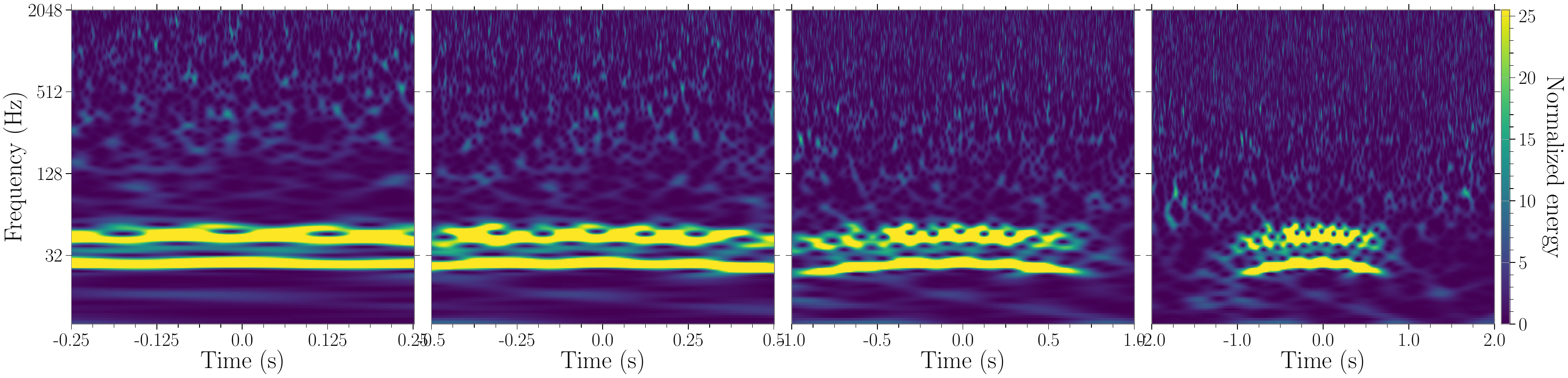}
         \caption[]{\small Example of Light Scattering glitch} 
         \label{fig:light-scattering}
     \end{subfigure}
     \hfill
     \begin{subfigure}[b]{\textwidth}
         \includegraphics[width=\textwidth]{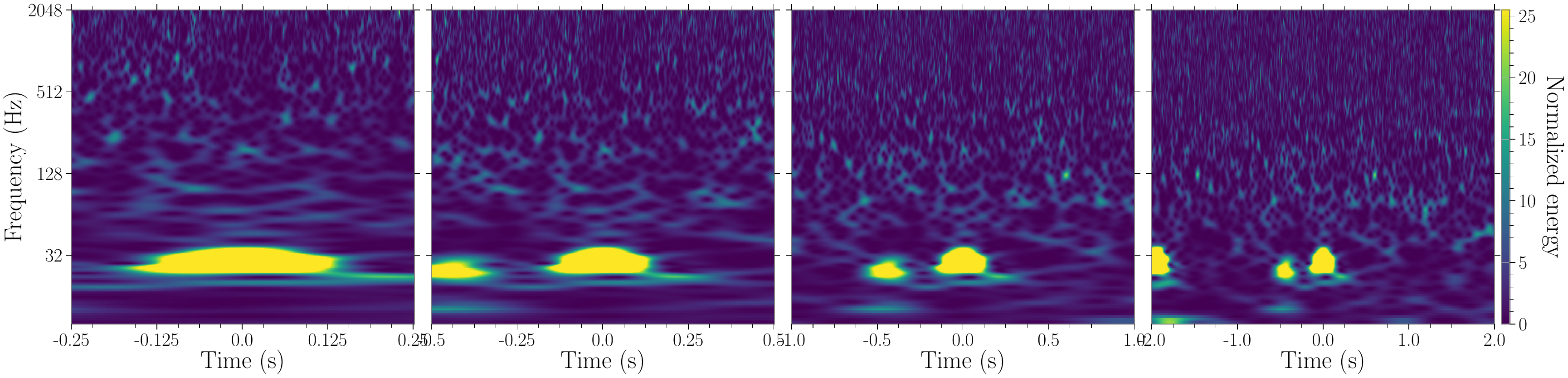}
         \caption[]{\small Example of Fast Scattering glitch} 
         \label{fig:fast-scattering}
     \end{subfigure}
     \hfill
     \begin{subfigure}[b]{\textwidth}
         \includegraphics[width=\textwidth]{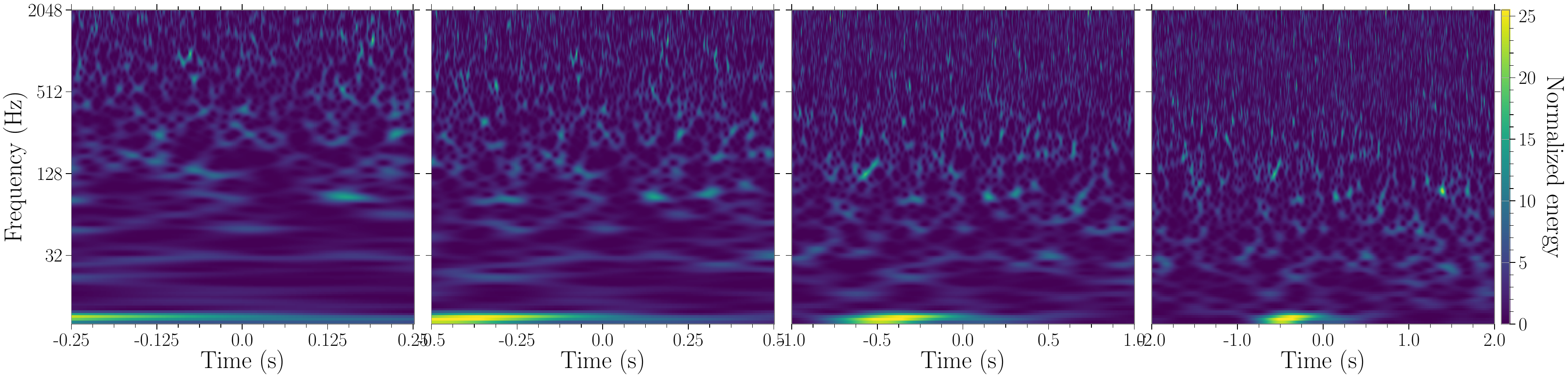}
         \caption[]{\small Example of Low-frequency Lines glitch} 
         \label{fig:lf-lines}
     \end{subfigure}
     \hfill
        \caption{Examples of the new glitch classes incorporated in the data set for \texttt{GSpyNetTree-O4}. The spectrograms are shown at four different durations (i.e., $0.5\;\mathrm{s}, 1\;\mathrm{s}, 2\;\mathrm{s},$ and $4\;\mathrm{s}$). We introduced these glitch classes as new classes to \texttt{GSpyNetTree-O4} due to their frequent recurrence in \ac{O3} and ER15. } 
        \label{fig:new-glitches}
\end{figure}

Table \ref{tab:class size} shows the total number of glitch samples included for each class, as well as the glitch classes considered by each classifier in \texttt{GSpyNetTree-O4}. In total, more than 80\% of the samples came from \ac{LHO} and \ac{LLO}, while the remainder were from Virgo. Compared to \texttt{GSpyNetTree}, the \ac{LM} classifier in \texttt{GSpyNetTree-O4} included Koi Fish glitches. While Koi Fish are not morphologically similar to the GWs in this mass range, loud and saturated glitches can nevertheless overlap with GWs, as was the case of GW170817~\cite{gw170817}. 
Additionally, we included Light Scattering and Fast Scattering as new classes in all classifiers since they were the most common glitch types during \ac{O3} and were responsible for broad ranges of vetted times~\cite{derek_validation}. We also introduced Low-frequency Lines as a new class in all classifiers since they were the most common glitch class during Engineering Run 15 (ER15)\footnote{ER15 was a pre-observing phase prior to \ac{O4}, which started on 2023-04-26.}. Figure~\ref{fig:new-glitches} shows examples of the three new glitch types included in \texttt{GSpyNetTree-O4}. 

\subsection{Construction of a GW set in representative background noise}\label{sec:noiseRobustness}

To generate \ac{GW} samples, we simulated signals and injected them into quiet (i.e., glitch-free) time series detector data. For \texttt{GSpyNetTree}, all quiet segments were manually selected from a small set of manually inspected 64-second time series data segments from \ac{LHO} and \ac{LLO}. However, the selected time series segments only covered a few days during the first (O1) and second (O2) observing runs~\cite{gspynettree1} and were not representative of the background noise of an entire observing run. Changes in detector configurations or environmental disturbances (e.g., earthquakes) may produce long-term variations in the background noise~\cite{LIGO:2024kkz}, which may be reflected in the spectrograms used as inputs to \texttt{GSpyNetTree-O4}. In fact, \textcite{gspynettree1} found that the performance of \texttt{GSpyNetTree} strongly depended on background noise.

Ensuring robustness to changes in background noise (i.e., steady-state, Gaussian detector noise) was a central focus of the \ac{O4} implementation of \texttt{GSpyNetTree}. For \texttt{GSpyNetTree-O4}, we expanded the range of quiet time series segments used as background noise. In addition to the quiet data segments from O1 and O2 used in \texttt{GSpyNetTree}, we implemented a method that automatically identified quiet \ac{O3} data segments into which we could inject simulated \ac{GW} signals. Our method for identifying quiet time series segments is similar to that of the \texttt{GlitchFind} \ac{DQR} task \cite{LIGO:2024kkz, Davis:2026kkz}. We calculated the energy of the time-frequency tiles of a spectrogram and verified Gaussianity by fitting an exponential distribution to the energy of the time-frequency tiles. If the normalized energy of four consecutive time-frequency bins exceeded 24, the time series segment corresponding to the spectrogram was flagged as a segment containing a glitch. This value was determined empirically. Otherwise, the time series segment was considered to be `quiet' and was used for injecting \ac{GW} signals. 

Using this procedure, we obtained 984, 980, and 958 distinct ten-second quiet time series segments from \ac{O3} for \ac{LHO}, \ac{LLO}, and Virgo, respectively. This considerably increased the number of quiet time series segments available. While the development of \texttt{GSpyNetTree-O4} was focused primarily on the implementation for LIGO detectors (\texttt{GSpyNetTree-O4} did not run on Virgo data), including quiet Virgo time series segments made \texttt{GSpyNetTree-O4} more robust to changes in background noise. Figures~\ref{fig:specs}c and~\ref{fig:specs}d show 10-second spectrograms of `quiet' time identified by the algorithm from \ac{LHO} and \ac{LLO}, respectively.

After quiet time series segments were identified, we simulated and injected \ac{GW} signals using a procedure similar to~\textcite{gspynettree1}. We first sampled parameters uniformly in the following ranges: total source-frame mass $M$ between $5\; M_\odot$ and $350\; M_\odot$, with individual component source-frame masses $m_1, m_2$ ranging from $2\; M_\odot$ to $175\; M_\odot$; \ac{SNR} between $8$ and $35$; individual dimensionless component spins $a_1, a_2$ between $0.05$ and $0.95$. We also randomized \ac{GW} source orientations and sky locations across the entire sky. This was an improvement over the data set for \texttt{GSpyNetTree}, which assumed the optimal orientation of \ac{GW} signals with respect to detectors. \ac{GW} signals were then simulated using \texttt{LALSuite} \cite{lalsuite} and the \texttt{IMRPhenomPv2} waveform model~\cite{waveform1, waveform2}. In total, we simulated 1800, 2100, and 1500 \acp{GW} for the \ac{LM}, \ac{HM}, and the \ac{EHM} classifiers, respectively, which were then injected into the time series obtained above. 

To mitigate overfitting, we applied the data augmentation techniques suggested by~\textcite{jarov}. For each simulated \ac{GW} signal, we generated four additional copies by randomly shifting the simulated \ac{GW} signal in time. The time offsets were drawn from a uniform distribution over $\pm0.5\;\mathrm{s}$ relative to the merger time. Having samples that were not always perfectly centered in the spectrograms made \texttt{GSpyNetTree-O4} robust to small offsets in candidate merger times, as first reported in \textcite{gspynettree1}. 

\subsection{Construction of an overlapping set of GWs and glitches}\label{sec:overlapping}

With increased detector sensitivity and a higher detection rate, the probability of \acp{GW} overlapping with or occurring in close proximity to glitches increases. This scenario already occurred with more than 20\% of candidates during \ac{O3}~\cite{o3a,o3b}, and more than 30\% of candidates in the first catalog released from \ac{O4} data~\cite{gwtc-4}. This issue has remained relevant throughout \ac{O4}~\cite{LIGO:2024kkz}. 

As described in \textcite{gspynettree1}, \texttt{GSpyNetTree} followed a multi-class architecture. This meant that, given an input, the classifiers could only predict a single class, which was assumed to be independent and mutually exclusive from the others. When tested on overlapping samples of \acp{GW} and glitches, \texttt{GSpyNetTree} classified more than 60\% (70\% for the \ac{HM} classifier) of \ac{GW} signals as only glitches~\cite{gspynettree1}. By identifying glitches as well as \acp{GW} in close proximity, \texttt{GSpyNetTree-O4} could also inform event retraction and enable timely decisions for critical detector-characterization tasks like glitch mitigation~\cite{derek, LIGO:2024kkz}.

To enable simultaneous identification of both glitches and \acp{GW} present in the same input, we generated samples of overlapping \acp{GW} and glitches for training and testing \texttt{GSpyNetTree-O4}. We also modified \texttt{GSpyNetTree-O4}'s classifier architecture to enable the simultaneous identification of glitches and \acp{GW}. In this section, we discuss the generation of overlapping samples. The architecture changes made to the classifiers are discussed in the next section.

\begin{figure}
\centering
\includegraphics[width=\linewidth]{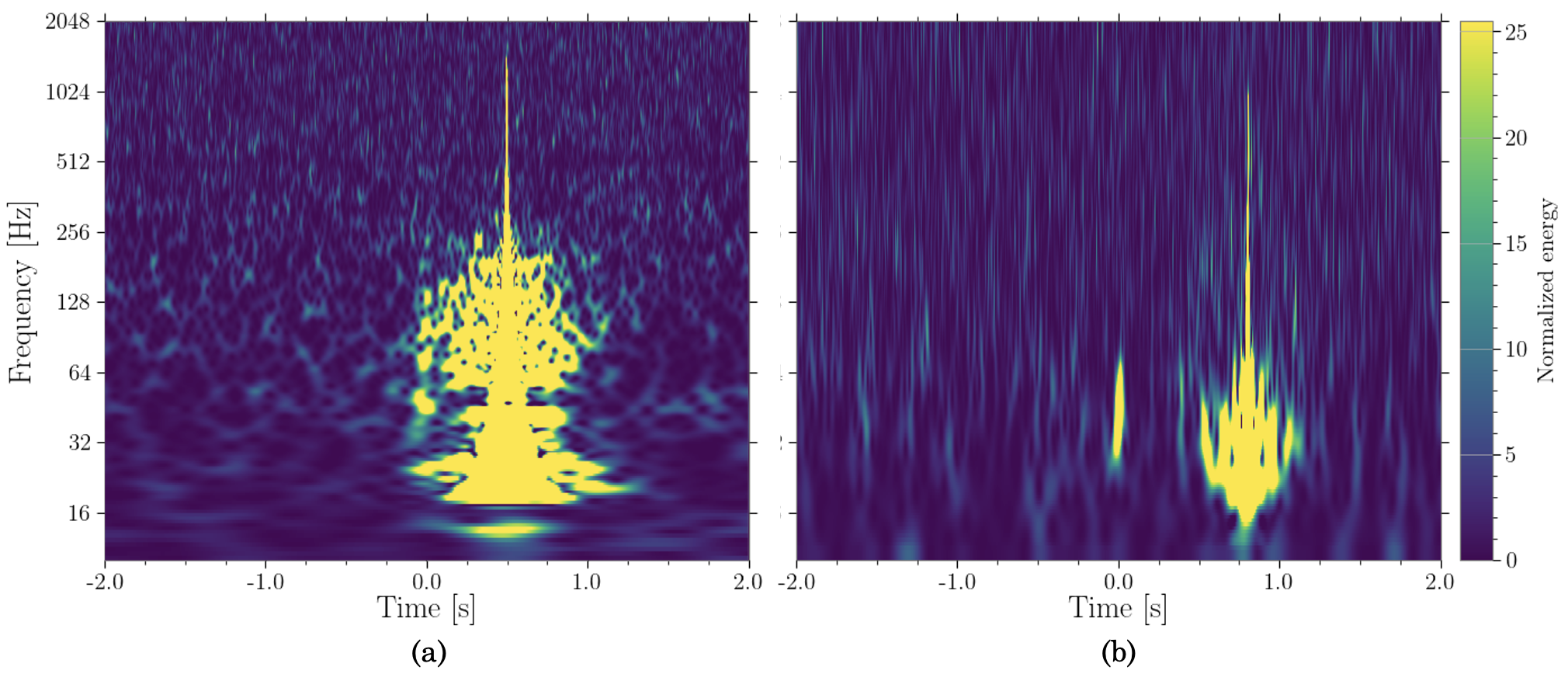}
\caption{(a) Example spectrogram of a Koi Fish glitch that fully obscures the injected \ac{GW} signal. The \ac{GW} is unidentifiable due to the excess power dominating the time-frequency visualization. For reference, we are interested in generating spectrograms like panel (b): a glitch (right) occurs in close proximity to a \ac{GW} signal (left), but without fully overlapping or obscuring it. }
\label{fig:saturated}
\end{figure}

To test the performance of \texttt{GSpyNetTree} on samples where \ac{GW} signals overlapped with a glitch,  \textcite{gspynettree1} generated 30 such samples for each glitch type. Each of them was created and manually verified by simulating 3 \ac{GW} signals, and shifting them with respect to the glitch, using 10 time offsets drawn from a Normal distribution with mean $\mu = 0\,\mathrm{s}$ and standard deviation $\sigma = 0.25\,\mathrm{s}$. With these parameters, a non-negligible number of glitches overlapped with the injected \acp{GW}. In some cases, this led to saturated spectrogram visualizations, where the \ac{GW} signal is completely obscured by the glitch. An example of this is shown in Figure \ref{fig:saturated}a, where a Koi Fish glitch completely overlapped with a simulated \ac{GW} signal. For reference, note that we were interested in generating examples like Figure \ref{fig:saturated}b: \acp{GW} and glitches (Koi Fish, in this case) in close proximity but not fully overlapping. In \textcite{gspynettree1}, the few fully overlapping samples could be replaced manually. However, this process became unmanageable for a larger training set. 

For \texttt{GSpyNetTree-O4}, we designed a new method to inject \ac{GW} signals near glitches. We selected 30 random examples of each glitch class and calculated their average duration, as shown in Table~\ref{tab:glitch durations}. We targeted the same common glitch classes as described in \textcite{gspynettree1}. We found that glitches related to light scattering (e.g., Scratchy, Fast Scattering, and Light Scattering glitches~\cite{sidScattering, 10.1063/5.0136896, LIGO:2024kkz}), were all long in duration ($>1.5\;\mathrm{s}$). However, compared to \acp{GW}, they occurred at rather low frequencies (except for Scratchy glitches, which in turn had lower \acp{SNR}). In contrast, glitches related to a sharp, step-function-like change in the time series data (e.g., Blips, Low-frequency Blips, Tomte, and Koi Fish glitches~\cite{PhysRevD.108.122004, LIGO:2024kkz}) were shorter in duration ($<1\;\mathrm{s}$) but could be found in a broader frequency range. 

To generate samples of overlapping \acp{GW} and glitches, we first determined whether the glitch was short (Blip, Low-frequency Blip, Tomte, or Koi Fish) or long (Fast Scattering, Scratchy, or Light Scattering). If short, we generated two samples per \ac{GW} signal: one sample with the glitch before the GW and one with the glitch after the GW signal. The offset in time between the \ac{GW} signal and each glitch was drawn from a uniform distribution. For each sample, we ensured that the offset was large enough such that the glitches were not added directly on top of the \ac{GW} signal. If the glitch was long, we generated three samples per \ac{GW} signal: in addition to the two samples described above, we added one glitch exactly at the \ac{GW} merger time. The reason we included this extra sample is twofold: first, it allowed us to represent in the data set the fact that long glitches like Light Scattering and Fast Scattering were the most common during \ac{O3}~\cite{derek,derek_validation}. Second, since they all occurred at low frequencies and/or had low \ac{SNR}, the probability that the resulting spectrogram would be over-saturated was low. 

\begin{table}[!htbp]
\centering
\small
\resizebox{0.5\textwidth}{!}{%
\begin{tabular}{lc}
\hline
\textbf{Glitch type} & \textbf{Mean duration [s]} \\ 
\hline
Blip & 0.138 \\
Low-frequency Blip & 0.137 \\
Tomte & 0.346 \\
Koi Fish & 0.899 \\
Fast Scattering & 1.657 \\
Scratchy & 1.750 \\
Light Scattering & 2.073 \\ 
\hline
\end{tabular}%
}
\caption{Average duration of the glitch types included in the training set for \texttt{GSpyNetTree-O4}. Note that Blips, Low-frequency Blips, Tomte, and Koi Fish glitches are shorter in duration ($<1\;\mathrm{s}$) than Fast Scattering, Scratchy, and Light Scattering glitches ($>1.6\;\mathrm{s}$).}
\label{tab:glitch durations}
\end{table}

To further minimize the probability of fully overlapping \acp{GW} and glitches, we standardized the $Q$-value of all our spectrograms to $Q=20$. For some samples, fixing the $Q$-value at $Q=20$ limited the energy per tile in our time-frequency visualizations, thereby lowering the likelihood of spectrogram saturation due to excess power. Additionally, it improved the visualization of low-mass \ac{GW} signals, increasing the probability that these signals were correctly identified. In Section~\ref{sec:q-value-val}, we tested the impact of samples generated using different $Q$-values on the performance of the classifiers. 

The lower section of Table~\ref{tab:class size} shows the samples of overlapping \acp{GW} and glitches for the \ac{LM}, \ac{HM}, and \ac{EHM} classifiers, respectively. Because of the short time interval between ER15 (which started on 2023-04-26) and \ac{O4} (which started on 2023-05-24), we were unable to generate overlapping samples of \acp{GW} and Low-frequency Lines. The \ac{GW} signals for all overlapping samples were generated with parameters drawn from the same ranges as described in Section~\ref{sec:noiseRobustness}. 

Additionally, in \ac{O4} the LIGO detectors implemented a nonlinear method to subtract alternating current (AC) $60\,\mathrm{Hz}$ power artifacts and harmonics, similar to the technique used for the \ac{O3} public data release~\cite{gwosc1}. Figure 5 in \textcite{gspynettree1} shows a Blip glitch with (right panel) and without (left panel) the nonlinear subtraction applied. While the differences in both spectrograms may appear to be negligible, \textcite{gspynettree1} showed that, when tested on glitches with the nonlinear subtraction applied, \texttt{GSpyNetTree}'s classification accuracy dropped from $>90\%$ to $\sim75\%$ for the three classifiers~\cite{gspynettree1}. To better reflect the data that was expected in \ac{O4}, we further augmented our data set by applying the nonlinear 60 Hz line-subtraction calibration to all our \ac{LHO} and \ac{LLO} glitches and \acp{GW}, including overlapping samples of \ac{GW} and glitches. This increased robustness to a broader range of background noise toward \ac{O4}. The total numbers of samples in our data set after the improvements and changes applied were $1.2\times10^5$, $1.3\times10^5$, and $1.0\times10^5$ for the \ac{LM}, \ac{HM} and \ac{EHM} classifiers, respectively. For each classifier, most of the samples were from the LIGO detectors, while approximately 20\% were from the Virgo detector. We used 80\% of the entire data set for training (20\% of which was used for validation). The remaining samples were used for testing. 

\subsection{Architecture}
For \texttt{GSpyNetTree-O4}'s three classifiers, we implemented a multi-label architecture, which allowed for the simultaneous identification of multiple classes within the same input~\cite {geron}. We kept the InceptionV3 \cite{inceptionv3} model from \texttt{GSpyNetTree} due to the high accuracy achieved in \textcite{gspynettree1}. To enable multi-label classification, we changed the output-layer activation function from \texttt{softmax} to \texttt{sigmoid} and optimized a binary cross-entropy loss for each output. To convert the output probabilities into predicted labels for evaluation, we treated a class as present when its probability exceeded 50\%. We monitored recall rather than single-label accuracy during training. Recall is defined as the ratio of true positives to the sum of true positives and false negatives. A false negative occurs when \texttt{GSpyNetTree-O4} fails to identify one or more objects present in the input; for example, when it detects no object in an input that contains one, or detects fewer than all objects in an input that contains multiple objects. By using recall as the evaluation metric, we directly quantified the fraction of glitches or \ac{GW} signals that the classifier identified. The change from multi-class architecture to multi-label architecture was the most significant upgrade we included in \ac{O4}, and it resulted in dramatic improvements in glitch identification in the presence of a \ac{GW} candidate. We discuss our results in the next section.

\section{Results}\label{sec:results}
The primary objective of \texttt{GSpyNetTree-O4} as a \ac{DQR} task was to identify \ac{DQ} issues: glitches in the vicinity of a \ac{GW} candidate. In this section, we evaluate the performance of \texttt{GSpyNetTree-O4} on the test set, which consisted of 11960, 13216, and 9891 samples for the \ac{LM}, \ac{HM}, and \ac{EHM} classifiers, respectively. Throughout the remainder of this paper, we used a threshold of 50\% to determine whether \texttt{GSpyNetTree-O4} predicted the presence of a given class in the input, consistent with the threshold used during training. Specifically, if the predicted probability for a class exceeded 50\%, \texttt{GSpyNetTree-O4} was considered to have identified that class. This enabled a direct comparison with the performance of \texttt{GSpyNetTree}. However, the DQR in O4 employed a more restrictive threshold for most statistical tasks, including \texttt{GSpyNetTree-O4}, following the framework described in~\textcite{Davis:2026kkz}. For \texttt{GSpyNetTree-O4}, this meant requiring a high predicted glitch probability before issuing a \ac{DQ} flag. This more restrictive criterion was expected to reduce the false alarm rate by requiring higher-confidence glitch predictions, but it could also lower the true-positive rate by missing lower-confidence glitches.

\begin{figure}[!htb]
\centering
\captionsetup[subfigure]{font=small}
     \begin{subfigure}[b]{0.32\textwidth}
         \includegraphics[width=\textwidth]{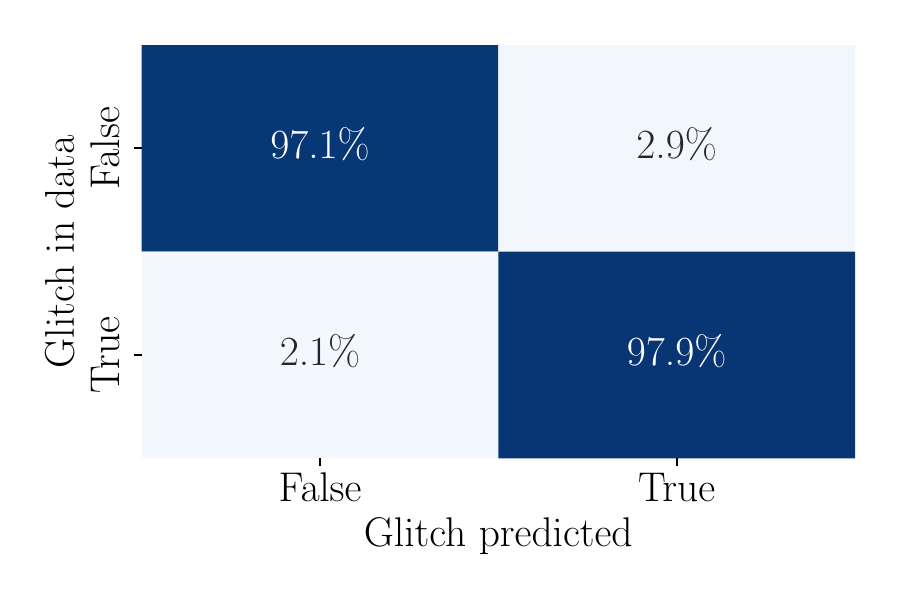}
         \caption[]{}
         \label{fig:lm binary}
     \end{subfigure}
     \hfill
     \begin{subfigure}[b]{0.32\textwidth}
         \includegraphics[width=\textwidth]{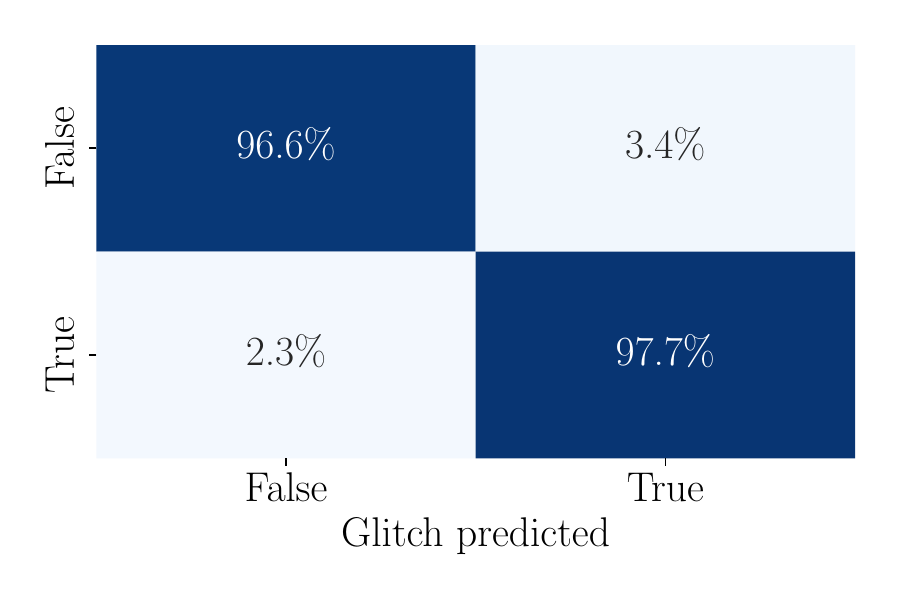}
         \caption[]{}
         \label{fig:hm binary}
     \end{subfigure}
     \hfill
     \begin{subfigure}[b]{0.32\textwidth}
         \includegraphics[width=\textwidth]{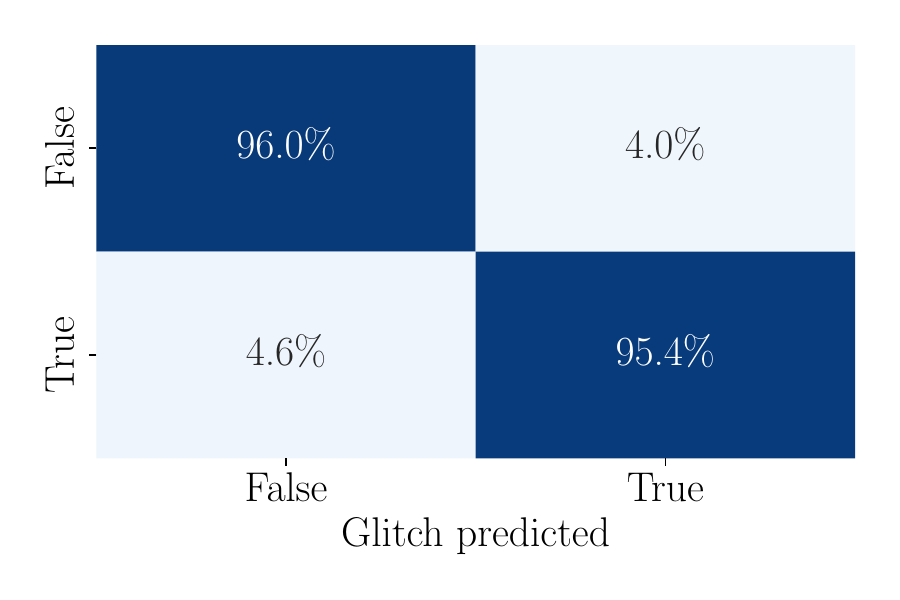}
         \caption[]{} 
         \label{fig:ehm binary}
     \end{subfigure}
     \hfill
        \caption{Glitch binary confusion matrices for the (a) \ac{LM}, (b) \ac{HM}, and (c) \ac{EHM} classifiers on the test data. The bottom-right cell represents correct glitch identifications by \texttt{GSpyNetTree-O4}. The classifiers identified 97.9\%, 97.7\%, and 95.4\% of glitches, respectively. The bottom-left cell corresponds to the false-negative rate (i.e., the percentage of glitches that were not identified by \texttt{GSpyNetTree-O4}), which is at most 4.6\%. The top-right cell corresponds to the false-positive glitch rate (i.e., the percentage of samples in which a glitch was incorrectly predicted by the classifiers), which is at most 4.0\%. Together, these matrices show high glitch-identification rates and low false-positive rates on the test sets.}
        \label{fig:binary-matrices}
\end{figure}

To assess \texttt{GSpyNetTree-O4}'s performance, we began by grouping all glitch classes under the same umbrella (including overlapping samples of \acp{GW} and glitches). This allowed us to determine whether a glitch was correctly flagged by \texttt{GSpyNetTree-O4}. Figure~\ref{fig:binary-matrices} presents binary confusion matrices for the \ac{LM} (left), \ac{HM} (center), and \ac{EHM} (right) classifiers, comparing the actual presence of glitches in the test data (vertical axis) with their predicted presence by \texttt{GSpyNetTree-O4} (horizontal axis). In general, we aimed to maximize the true positive rate, shown in the bottom-right entry of each confusion matrix. The true-positive rate represents the fraction of glitches that were correctly identified by \texttt{GSpyNetTree-O4}. We obtained true-positive rates of 97.9\%, 97.7\%, and 95.4\% for the \ac{LM}, \ac{HM}, and \ac{EHM} classifiers, respectively. These values corresponded to low false-negative rates for glitches, as shown by the bottom-left entries of the confusion matrices. While the \ac{EHM} classifier had the lowest true-positive rate, it was also the least frequently used, as it was applied only when the candidate event's mass exceeded $250\; M_\odot$. Through O4b, the most massive candidate detected by the \ac{LVK} had a total mass of $M=236^{+29}_{-48}\;M_\odot$~\cite{gwtc-4}. This event was handled by \texttt{GSpyNetTree-O4}'s \ac{HM} classifier.

We were also interested in achieving a low false-positive rate, shown in the top-right entry of each confusion matrix. A high false-positive rate would have meant that \ac{GW} candidates were frequently and incorrectly flagged as having \ac{DQ} issues. This would have triggered unnecessary additional event validation and reduced the trustworthiness of \texttt{GSpyNetTree-O4}. The false-positive glitch rates were 2.9\%, 3.4\%, and 4.0\% for the \ac{LM}, \ac{HM}, and \ac{EHM} classifiers, respectively. Equivalently, the classifiers achieved high true-negative rates, shown in the top-left entries of the confusion matrices. This means that, among samples with no glitches present in the data, 97.1\%, 96.6\%, and 96.0\%, respectively, were correctly classified as having no \ac{DQ} issue.

\begin{figure}[!htb]
\centering
\captionsetup[subfigure]{font=small}
     \begin{subfigure}[b]{0.49\textwidth}
         \includegraphics[width=\textwidth]{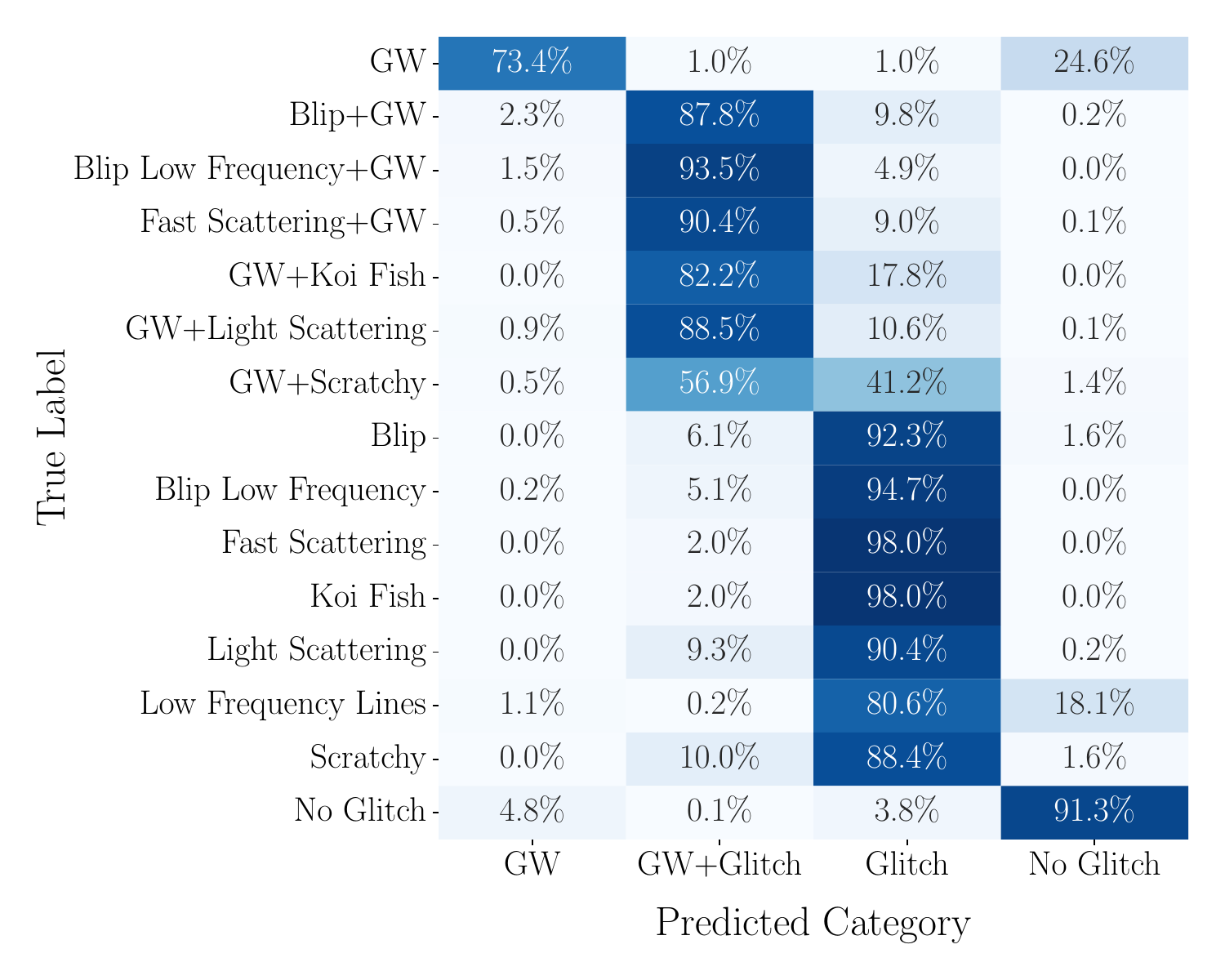}
         \caption[]{} 
         \label{fig:specific-lm}
     \end{subfigure}
     \hfill
     \begin{subfigure}[b]{0.49\textwidth}
         \includegraphics[width=\textwidth]{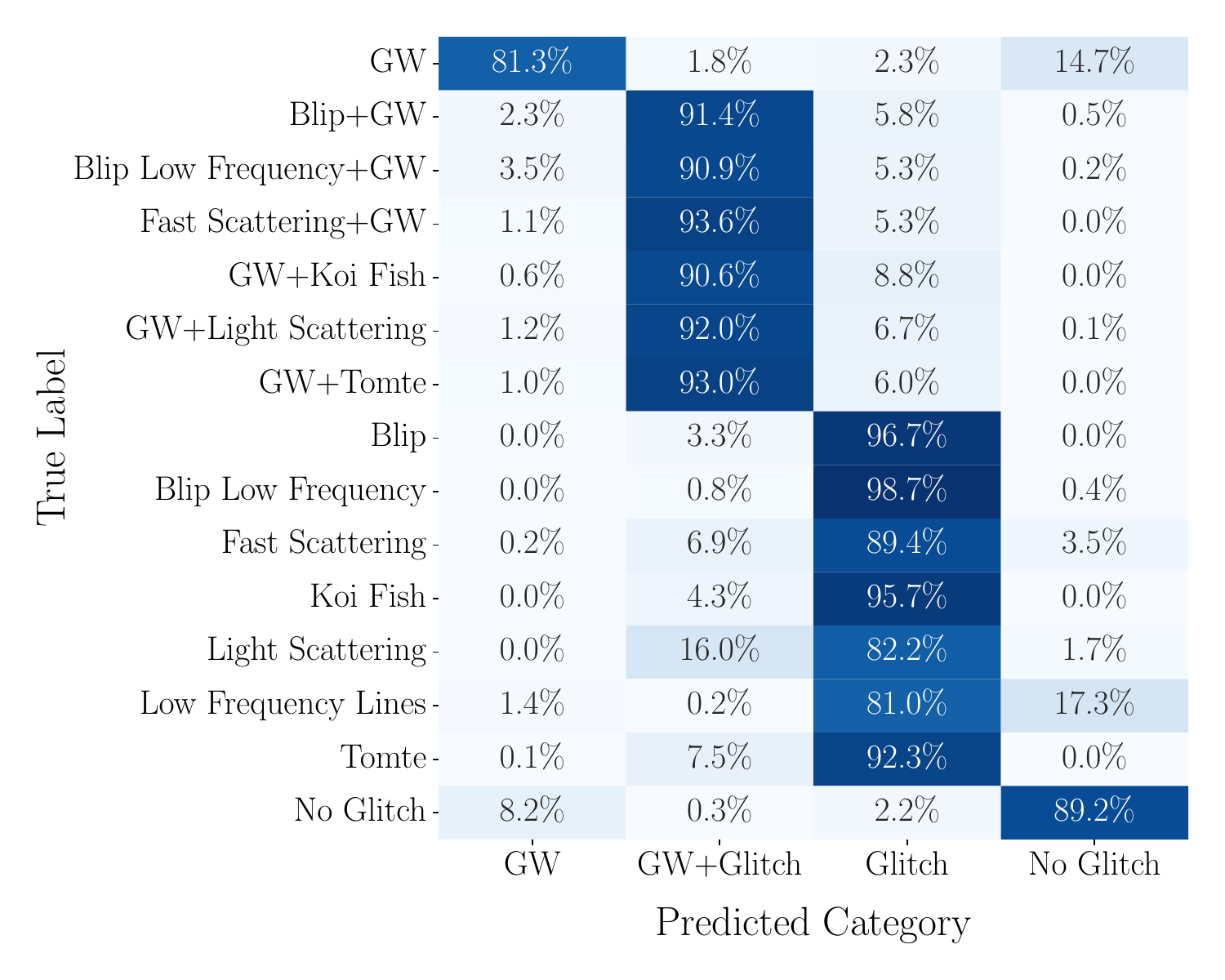}
         \caption[]{} 
         \label{fig:specific-hm}
     \end{subfigure}

     \vskip\baselineskip
     \begin{subfigure}[b]{0.48\textwidth}
         \includegraphics[width=\textwidth]{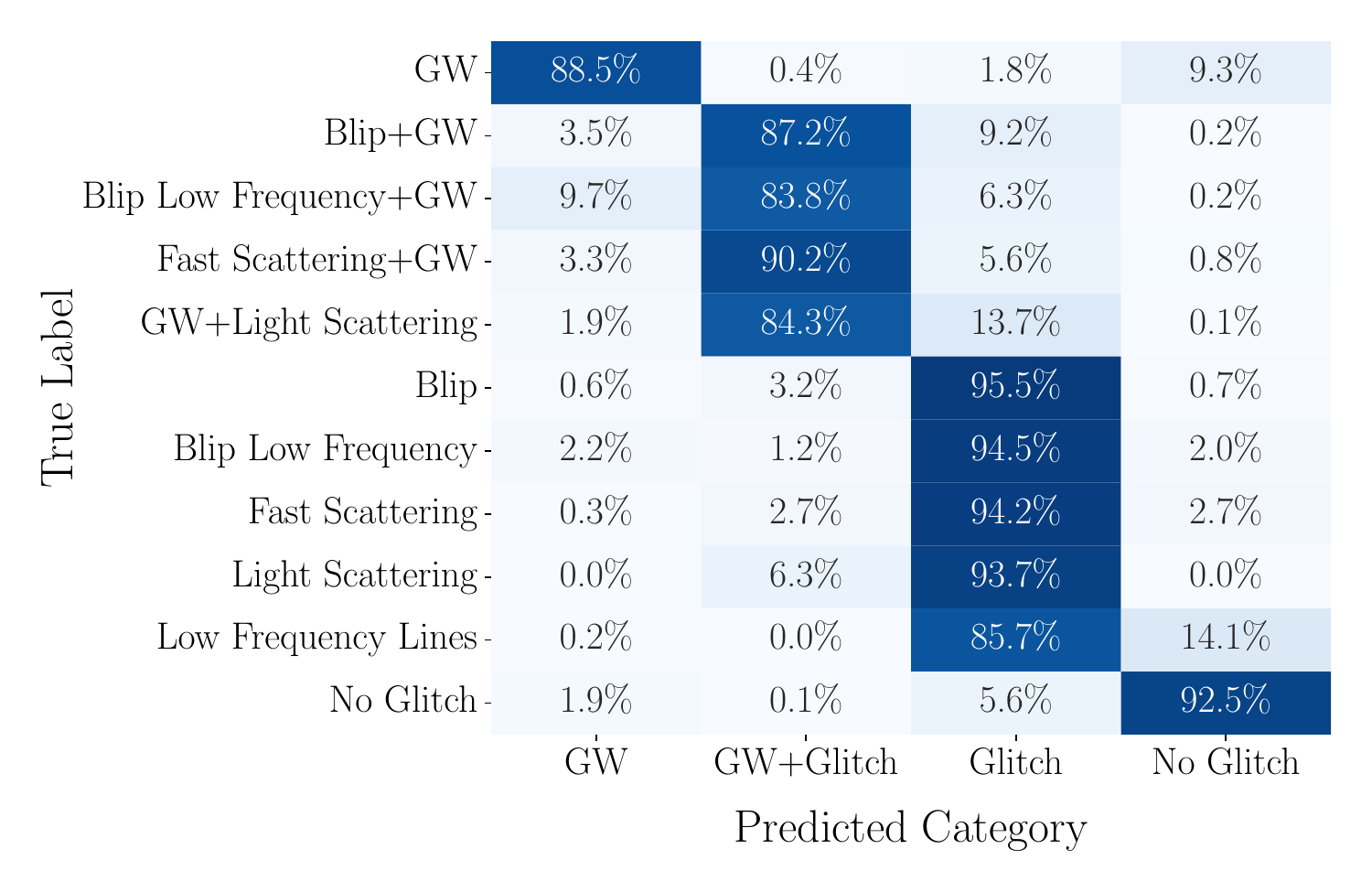}
         \caption[]{}
         \label{fig:specific-ehm}
     \end{subfigure}
     \hfill
        \caption{Specific confusion matrices for the (a) \ac{LM}, (b) \ac{HM}, and (c) \ac{EHM} classifiers on the test data. The predicted labels on the $x$-axis are grouped into four categories: No Glitch, \ac{GW}, Glitch (including all relevant categories), and \ac{GW}+Glitch (samples in which both a \ac{GW} and a glitch were predicted). The $y$-axis shows all true labels of the test samples. Each row is normalized to 100\%. For example, $92.3\%$ of Blips were classified by the \ac{LM} classifier in the Glitch category. }
        \label{fig:specific-matrices}
\end{figure}

To further understand the results of \texttt{GSpyNetTree-O4} based on the labels on which it was trained, we grouped the predicted labels into four broad categories: \ac{GW}, Glitch, both (\ac{GW}$+$Glitch), or neither (No Glitch). We compared these to the true labels. The results are shown in Figure~\ref{fig:specific-matrices} as confusion matrices for each classifier. In general, the results of the three classifiers were very robust, consistent with the high rate of correct glitch identification shown in Figure~\ref{fig:binary-matrices}.

The first row of each confusion matrix in Figure~\ref{fig:specific-matrices} shows that 73.4\%, 81.3\%, and 88.5\% of the simulated \ac{GW} signals were correctly classified by the \ac{LM}, \ac{HM}, and \ac{EHM} classifiers, respectively. A substantial fraction of low-\ac{SNR} signals were instead classified as No Glitch: 24.6\%, 14.7\%, and 9.3\% for the \ac{LM}, \ac{HM}, and \ac{EHM} classifiers, respectively. This effect was most pronounced for the \ac{LM} classifier because low-\ac{SNR} signals in this mass range have more extended time-frequency morphologies and therefore appear fainter than signals with the same \ac{SNR} in the \ac{HM} and \ac{EHM} mass ranges. This makes them more difficult to distinguish from the No Glitch class. Ideally, in the presence of a real \ac{GW} event, the \ac{GW} label would be predicted instead of the No Glitch label. However, for event validation purposes, a \ac{GW} classified as a No Glitch will not trigger further event validation. Combining the results of the \ac{GW} and No Glitch prediction categories, we find that the \ac{LM}, \ac{HM}, and \ac{EHM} classifiers correctly flagged no \ac{DQ} issues in 98.0\%, 96.0\%, and 97.8\% of these samples, respectively. Equivalently, they falsely predicted a glitch in only 2.0\%, 4.0\%, and 2.2\% of the cases.

We next consider the rows of Figure~\ref{fig:specific-matrices} corresponding to single-glitch test samples. More than 88\%, 82\%, and 93\% of the samples from every glitch class except Low-frequency Lines were classified as Glitch only by the \ac{LM}, \ac{HM}, and \ac{EHM} classifiers, respectively. Counting both Glitch and \ac{GW}$+$Glitch predictions as successful glitch detections, these percentages increased to more than 98\%, 96\%, and 95\%. Although some samples also received a spurious \ac{GW} label, our main goal for \texttt{GSpyNetTree-O4} was to identify the presence of glitches. These high glitch-identification rates highlight the model's ability to flag \ac{DQ} issues.

However, the fraction of correctly identified glitches was notably lower for the Low-frequency Lines class. Only 80.8\%, 81.2\%, and 85.7\% of test samples from this class were correctly identified by the \ac{LM}, \ac{HM}, and \ac{EHM} classifiers, respectively, while $18.1\%$, $17.3\%$, and $14.1\%$ were misclassified as No Glitch. This is a consequence of the fact that we did not generate overlapping samples of Low-frequency Lines and \acp{GW} due to the short time between ER15 and \ac{O4}. As a result, Low-frequency Lines were underrepresented in the training set. We expect the performance for this class to improve in future observing runs after adding Low-frequency Lines samples overlapping with \ac{GW} signals. The lower fraction of correctly identified Low-frequency Lines samples, compared to the other glitch classes, highlights the importance of balanced and robust training sets in machine-learning applications.

\subsection{Overlapping GWs and glitches}
We now examine the rows of Figure~\ref{fig:specific-matrices} corresponding to samples containing overlapping glitches and \ac{GW}
signals. We first evaluated whether \texttt{GSpyNetTree-O4} correctly identified both components present in each sample: the glitch and the \ac{GW} signal. The \ac{HM} classifier performed best under this criterion: for all samples with overlapping \acp{GW} and glitches, \texttt{GSpyNetTree-O4} correctly identified both components in more than 90\% of the cases. The \ac{LM} classifier correctly identified both components in $\geq 87\%$ of the samples, except for two scenarios: \ac{GW}+Koi Fish (82.2\%) and \ac{GW}+Scratchy (56.9\%). Since both Koi Fish and Scratchy span a broad frequency range and are typically very loud, \acp{GW} of moderate to low \acp{SNR} tend to be buried in their presence. Consistent with this interpretation, 17.8\% of \ac{GW}+Koi Fish samples were classified as Koi Fish only and 41.2\% of \ac{GW}+Scratchy samples were classified as Scratchy only. 

\begin{figure}[htbp]
\centering
\captionsetup[subfigure]{font=small}
     \begin{subfigure}[b]{\textwidth}
         \includegraphics[width=\textwidth]{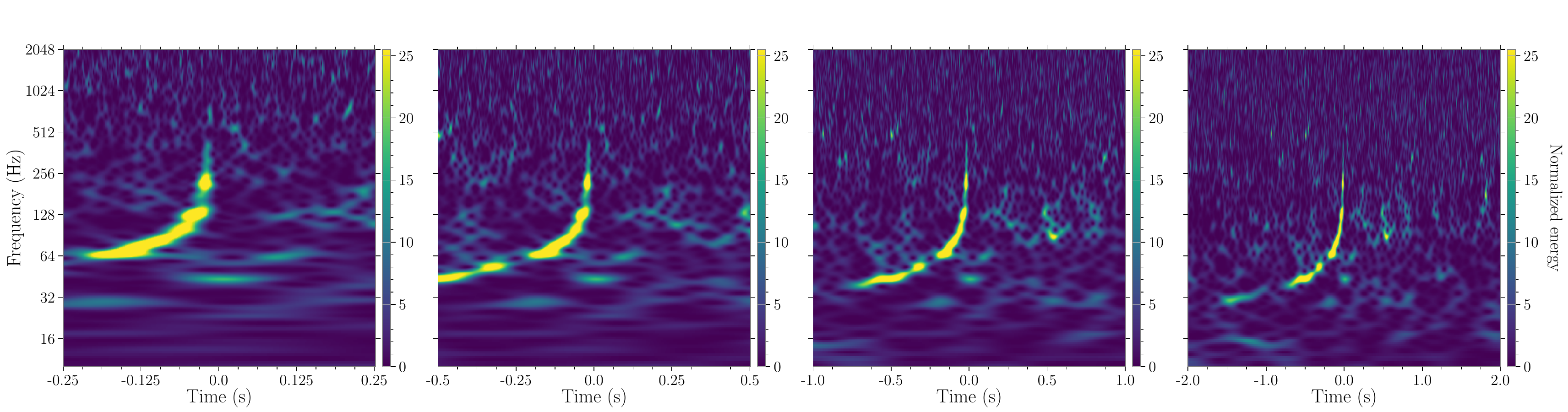}
         \caption[]{} 
         \label{fig:scratchy-gw-not-faint}
     \end{subfigure}
     \hfill
     \begin{subfigure}[b]{\textwidth}
         \includegraphics[width=\textwidth]{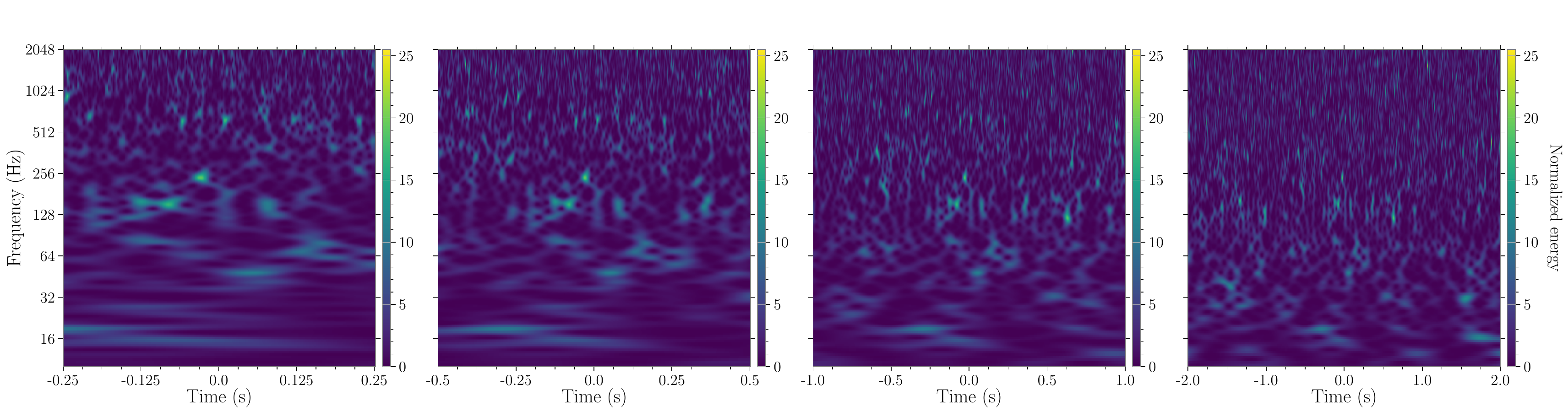}
         \caption[]{} 
         \label{fig:scratchy-gw-faint}
     \end{subfigure}
     \hfill
        \caption{Samples of Scratchy glitches overlapping with \ac{GW} signals with total masses (a) $M=23.7\;M_\odot$ and (b) $M=10.9\;M_\odot$. While the glitch is noticeable in both samples, the \ac{GW} is only visible to the naked eye in the top spectrogram, since it has higher total mass and SNR. In fact, \texttt{GSpyNetTree-O4} detected the glitch in both cases, but could only identify the GW signal in the top panel. } 
        \label{fig:scratchy-gw}
\end{figure}

Figure~\ref{fig:scratchy-gw} shows two examples of Scratchy glitches overlapping with \ac{GW} signals with total masses $M=23.7\;M_\odot$ (top panel) and $M=10.9\;M_\odot$ (bottom panel). In the first case, both glitch and GW were correctly identified, whereas in the second case only the glitch was detected. Because the primary purpose of \texttt{GSpyNetTree-O4} was to flag the presence of glitches, its performance on these overlapping samples is more appropriately summarized by the glitch identification rate (i.e., the glitch recall). For \ac{GW}+Koi Fish and \ac{GW}+Scratchy samples, \texttt{GSpyNetTree-O4} identified the glitch component in 100\% and 98.1\% of cases, respectively. 

\begin{figure}
\centering
\includegraphics[width=\linewidth]{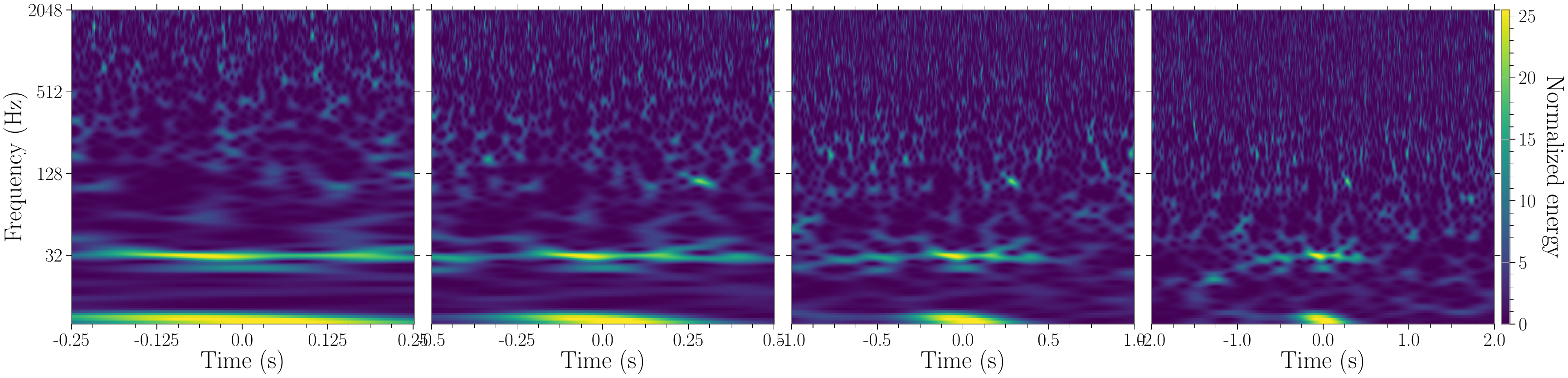}
\caption{Spectrogram showing a sample in which \texttt{GSpyNetTree-O4} predicts both a Light Scattering glitch at $f\approx 32\;\mathrm{Hz}$ and a Low-frequency Line at $f\approx 12\;\mathrm{Hz}$. }
\label{fig:overlapping-ehm-specs}
\end{figure}

Due to its multi-label nature, \texttt{GSpyNetTree-O4} could also predict more than two labels for a single input. Besides flagging a single glitch and a \ac{GW} signal, it could also predict multiple glitch classes simultaneously. For instance, the time-frequency visualizations in Figure~\ref{fig:overlapping-ehm-specs} show a Light Scattering glitch overlapping with a Low-frequency Line. \texttt{GSpyNetTree-O4} successfully detected both glitches, demonstrating its robustness in flagging DQ issues. 

\begin{figure}[!htb]
\centering
\captionsetup[subfigure]{font=small}
    \begin{subfigure}[b]{0.49\textwidth}
    \includegraphics[width=\textwidth]{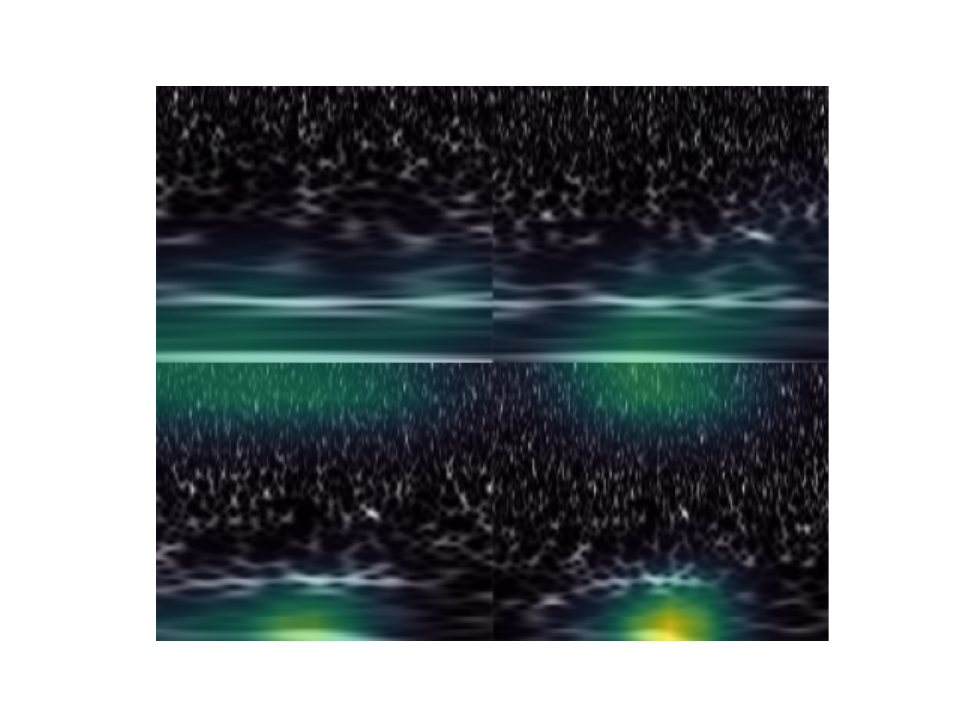}
    \caption[]{}
    \label{fig:tf-explain-ehm-intermediate}
\end{subfigure}%
\begin{subfigure}[b]{0.49\textwidth}
    \includegraphics[width=\textwidth]{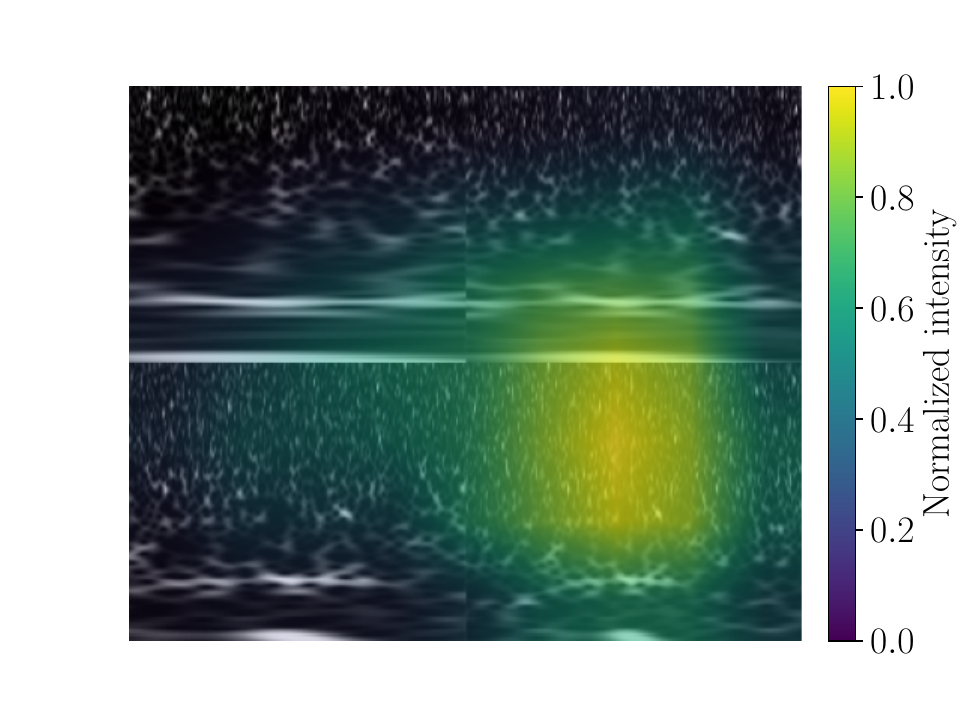}
    \caption[]{}
    \label{fig:tf-explain-ehm-final}
\end{subfigure}
\caption{Heatmaps generated using the \ac{Grad-CAM}~\cite{gradcam} module of \texttt{tf-explain}~\cite{tf-explain} for (a) an intermediate and (b) final convolutional layer of the \texttt{GSpyNetTree} InceptionV3 architecture, applied to the sample in Figure~\ref{fig:overlapping-ehm-specs}. Brighter areas indicate higher relative intensity, on a scale from 0 to 1. It can be seen that in the intermediate layers, the CNN focuses on the Low-frequency Line in both spectrograms, while in the final convolutional layer, it focuses on broader features containing the Light Scattering glitch.}
\label{fig:tf-explain}
\end{figure}

To understand how both glitches were identified, we use \texttt{tf-explain}~\cite{tf-explain}, a \texttt{TensorFlow} package to interpret neural network output. In particular, the \ac{Grad-CAM} module generates heatmaps that highlight regions of the $2\times2$ input spectrogram matrix contributing most to the \ac{HM} classifier's decision \cite{gradcam}. Figure~\ref{fig:tf-explain-ehm-intermediate} shows such a heatmap from an intermediate layer of our InceptionV3 architecture \cite{inceptionv3}, while Figure~\ref{fig:tf-explain-ehm-final} presents the one from its final convolutional layer, both computed for the sample in Figure~\ref{fig:overlapping-ehm-specs}. The yellow regions in the heatmap indicate areas of higher focus for the \ac{HM} classifier. As seen in the bottom spectrograms of the $2\times2$ matrix in Figure~\ref{fig:tf-explain-ehm-intermediate}, the intermediate layers of the \ac{HM} classifier focus on low-frequency regions, corresponding to the Low-frequency Line. As the input is processed through deeper layers of the network, the attention shifts to other regions, highlighting features associated with the Light Scattering glitch, as shown in Figure~\ref{fig:tf-explain-ehm-final}. We observed this same behavior in other overlapping samples of \acp{GW} and glitches in the test set. This suggests that the network is learning to identify overlapping features at different depths of the network, gradually building up a more complete understanding of the input spectrograms, and highlighting the relevance of a robust architecture like InceptionV3.

\subsection{An example of \texttt{GSpyNetTree-O4}'s performance on a retracted event}

We use the retracted candidate S230708bi to illustrate how \texttt{GSpyNetTree-O4} contributed to \ac{LVK} event validation in practice. This candidate was identified in low latency during \ac{O4a} on July 8 2023 with a false alarm rate of one per 0.44 years and an estimated total mass $\gtrsim 50\,M_\odot$. It was subsequently circulated via NASA's Gamma-ray Coordinates Network~\cite{GCN_notice_S230708bi} as a binary black hole merger candidate.

\begin{figure}
\centering
\includegraphics[
    width=0.85\linewidth,
    trim={2cm 3.5cm 2cm 2cm},
    clip
]{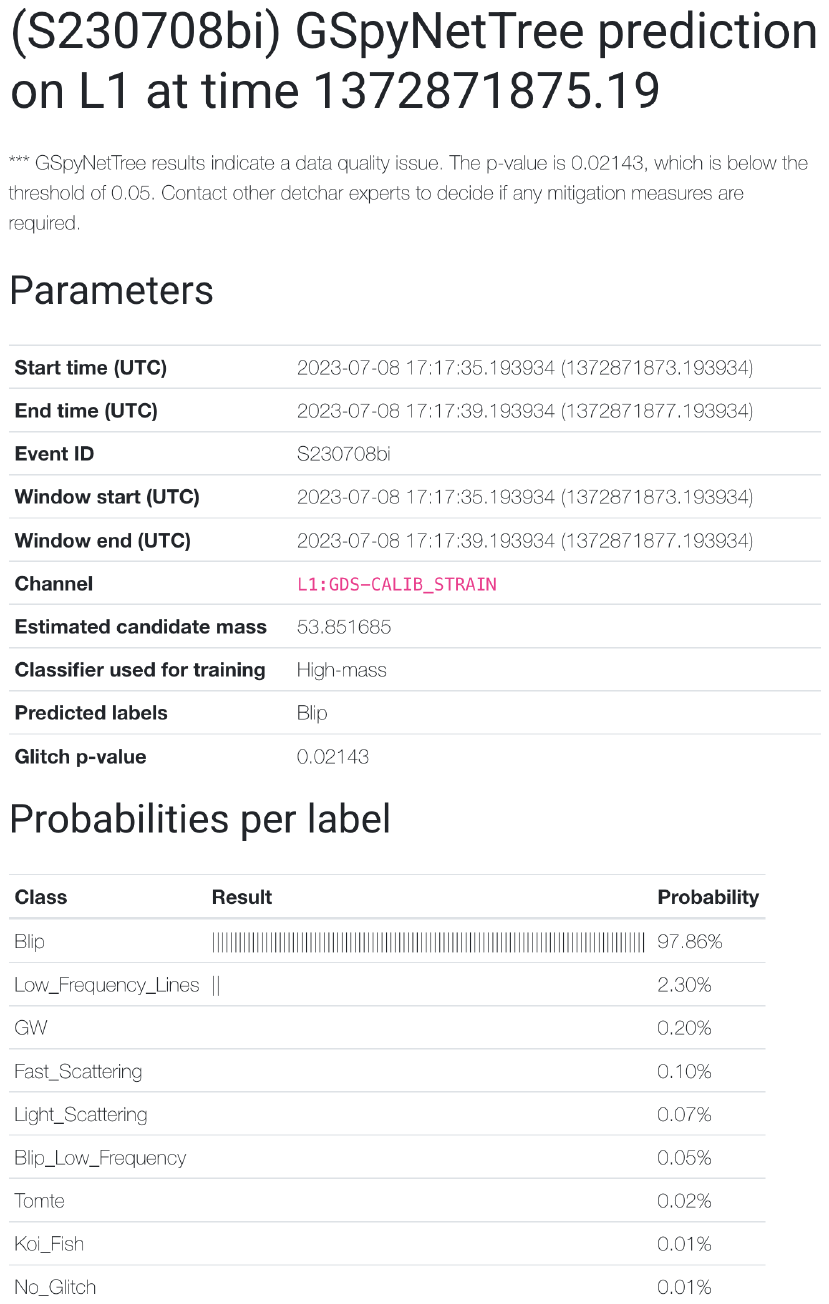}
\caption{Excerpt from the summary page generated by \texttt{GSpyNetTree-O4} in the DQR for the retracted candidate S230708bi at \ac{LLO}. The candidate was classified by \texttt{GSpyNetTree-O4} as a Blip glitch with a probability of 97.8\%. A data-quality issue was flagged as a result.}
\label{fig:page}
\end{figure}

\begin{figure}
\centering
\includegraphics[width=\linewidth]{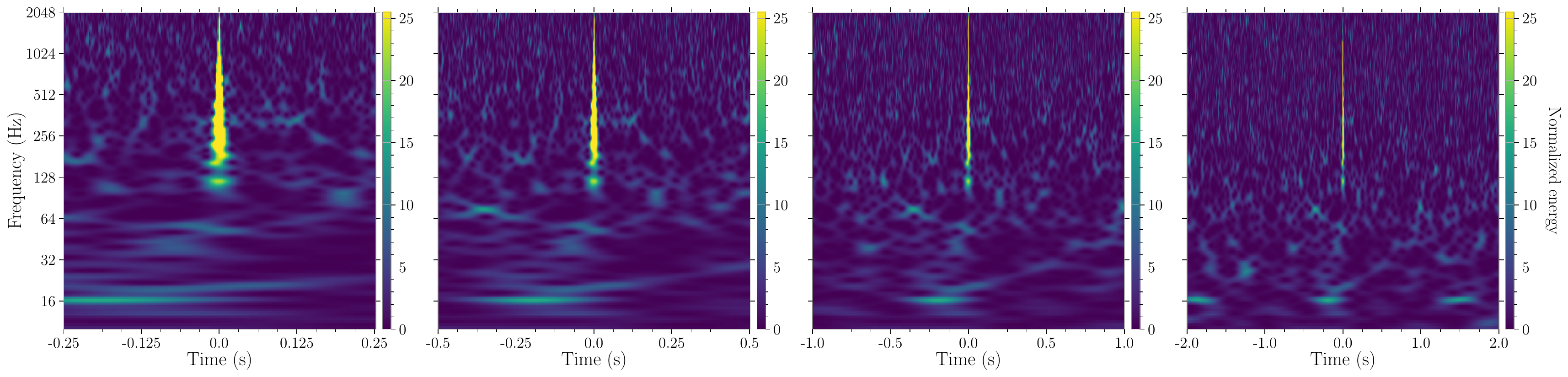}
\caption{Spectrograms of \ac{LLO} data surrounding a \ac{GW} candidate during \ac{O4a}. The candidate, S230708bi, was initially estimated to be a binary black hole merger and a \ac{GW} alert was publicly issued, but was later retracted. \texttt{GSpyNetTree-O4} correctly identified a Blip glitch at the estimated merger time, contributing to the retraction decision.}
\label{fig:S230708bi}
\end{figure}

In parallel, the candidate triggered the \ac{HM} classifier of \texttt{GSpyNetTree-O4}, which correctly identified a Blip at the estimated merger time. The candidate was retracted approximately 40 minutes after its initial identification. \texttt{GSpyNetTree-O4} contributed significantly to the data quality assessment that informed the retraction of this candidate~\cite{GCN_S230708bi}. In Figure~\ref{fig:page}, we show an excerpt of the summary page generated by \texttt{GSpyNetTree-O4} in the \ac{DQR} for this candidate. \texttt{GSpyNetTree-O4} identified the glitch with a probability of 97.8\%, flagging a DQ issue. The spectrograms used by \texttt{GSpyNetTree-O4} for this candidate are shown in Figure~\ref{fig:S230708bi}.

\section{Further studies on \texttt{GSpyNetTree-O4} for future observing runs}\label{sec:furtherstudies}

We perform a series of validation studies to assess \texttt{GSpyNetTree-O4}'s ability to generalize beyond the glitch morphologies represented in its training set. We suggest new avenues to improve glitch identification as detector sensitivity and candidate rates increase. We start by testing \texttt{GSpyNetTree-O4} on \texttt{Gravity Spy} glitch types we did not include in our training set as a proxy for evaluating \texttt{GSpyNetTree-O4}'s performance on glitches with new and unknown morphologies. We then test the effect of $Q$-value for spectrogram generation on glitch and \ac{GW} classification. Finally, we evaluate \texttt{GSpyNetTree-O4}'s performance on a set of O4 Virgo glitches.

\subsection{\texttt{GSpyNetTree-O4}'s performance on unknown glitch types}\label{sec:new-glitches-val}

\begin{figure}
\centering
\includegraphics[width=0.8\linewidth]{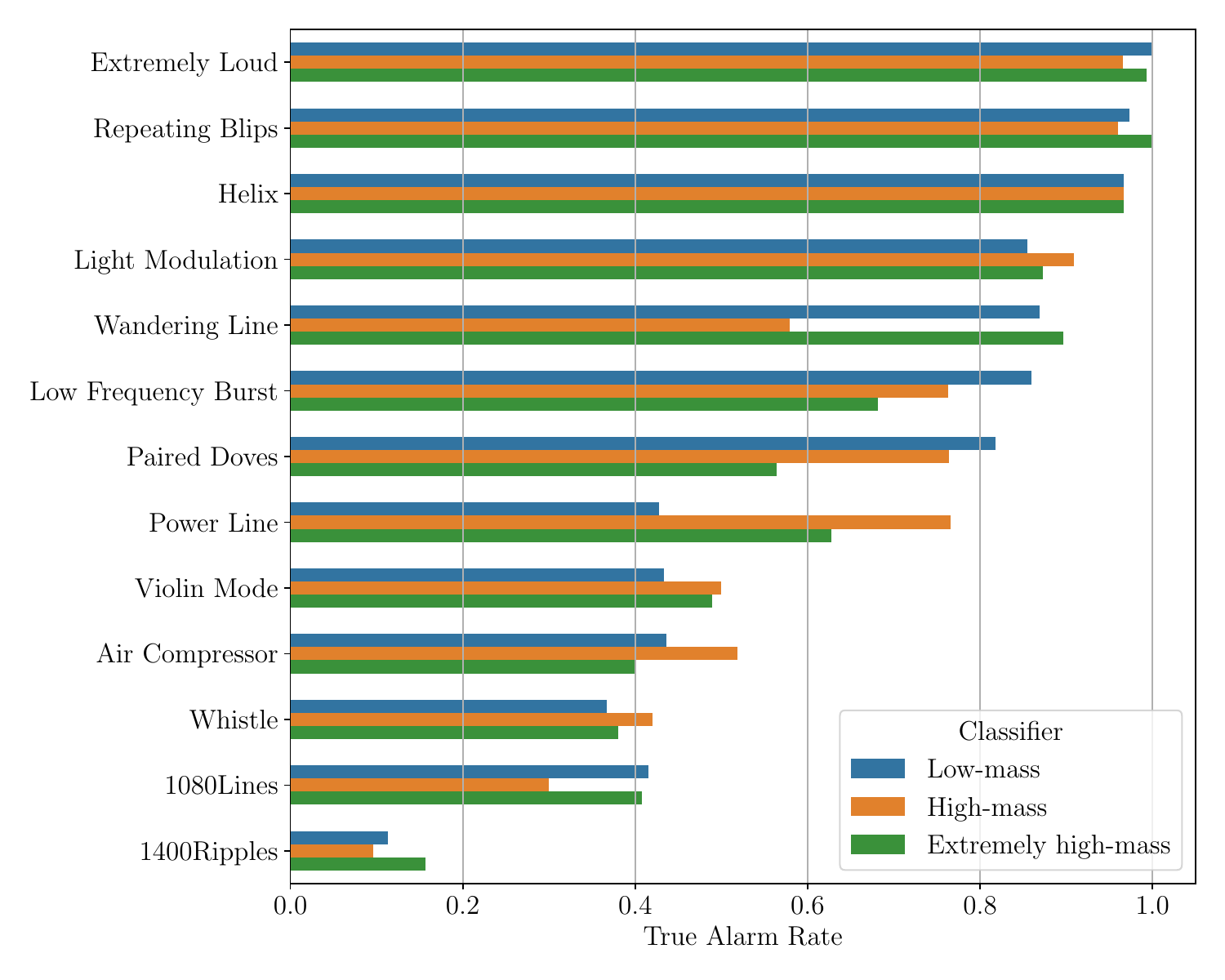}
\caption{Probability that a sample from the \texttt{Gravity Spy} glitches not included in \texttt{GSpyNetTree-O4}'s training set is classified as being/containing a glitch for the \ac{LM} (blue), \ac{HM} (orange), and \ac{EHM} (green) classifiers. }
\label{fig:other-glitches}
\end{figure}

\begin{figure}
\centering
\includegraphics[width=\linewidth]{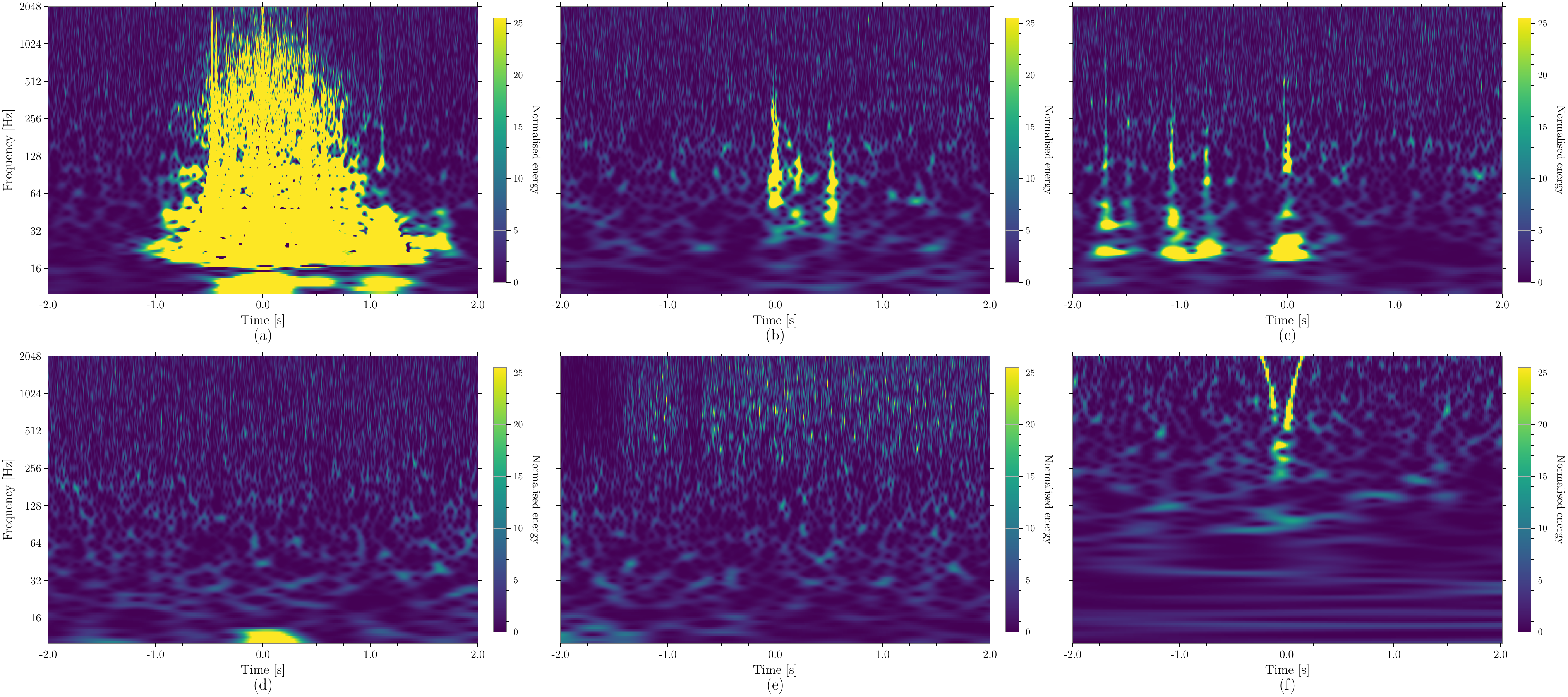}
\caption{Time-frequency visualizations of some glitches not included in \texttt{GSpyNetTree-O4}'s original training set, used for the validation study described in Section~\ref{sec:new-glitches-val}. The example glitches are: (a) Extremely Loud, (b) Repeating Blips, (c) Helix, (d) Low-frequency Burst, (e) Wandering Line, (f) Whistle. }
\label{fig:other-spectrograms}
\end{figure}

We tested \texttt{GSpyNetTree-O4}'s performance on all types of glitches classified by \texttt{Gravity Spy}~\cite{jane} that we did not include in the data set described in Section~\ref{sec:methods}. This test assessed whether \texttt{GSpyNetTree-O4} could correctly classify inputs with morphologies different from those represented in the training set. To do this, we first selected approximately 30 glitches per class that were confidently classified by \texttt{Gravity Spy} (i.e., with \texttt{Gravity Spy} confidence greater than $90\%$). We then ran \texttt{GSpyNetTree-O4} on these glitches. Figure~\ref{fig:other-glitches} shows the true positive rate for each of the tested glitch types for all three classifiers. The true positive rate for a glitch type is defined as the fraction of tested samples of that glitch type that were correctly classified as glitches. For the Extremely Loud samples, the three classifiers achieved an average true positive rate $\geq 98\%$. This was expected since this glitch is morphologically similar to Koi Fish: both display saturated spectrograms, as shown in Figure~\ref{fig:other-spectrograms}a. We found a similar behavior with \texttt{GSpyNetTree}, but only for the \ac{HM} classifier, which was the only one trained on Koi Fish glitches. The inclusion of Koi Fish glitches in the training set likely allowed \texttt{GSpyNetTree-O4} to generalize to similar saturated glitches in the LIGO detectors across the three classifier mass ranges.

Additionally, for Repeating Blips (see Figure~\ref{fig:other-spectrograms}b), we obtained true alarm rates of 97\%, 96\%, and 100\% for the \ac{LM}, \ac{HM}, and \ac{EHM} classifiers, respectively. This represented a significant improvement over \texttt{GSpyNetTree}, where only $\sim71\%$ (\ac{LM}), $\sim85\%$ (\ac{HM}), and $\sim53\%$ (\ac{EHM}) of the Repeating Blips were classified as glitches. Many samples of Repeating Blips were classified by \texttt{GSpyNetTree-O4} as a Blip (or other glitch) together with a non-existent \ac{GW}. This suggested that the improved classification of Repeating Blips was partially a result of the multi-label architecture and the inclusion of overlapping samples in the training set. Together, these enabled \texttt{GSpyNetTree-O4} to assign one instance of a blip as a Blip while assigning the others to different classes. Since in \texttt{GSpyNetTree-O4} we were only interested in glitch identification, these classifications were acceptable. One potential solution is to deploy a segmentation model such as \texttt{GSpyNetTree-S}~\cite{Chan:2025zab}, which can identify and localize multiple instances of one or more classes in a single input.

Moreover, \texttt{GSpyNetTree-O4} correctly identified $\sim97\%$ of Helix samples (Figure~\ref{fig:other-spectrograms}c) and $\sim87\%$ of Light Modulation samples. These glitch types are morphologically similar to classes included in the \texttt{GSpyNetTree-O4} training set. These results demonstrate \texttt{GSpyNetTree-O4}'s ability to generalize predictions to similar, but new glitch morphologies. This was also the case for Low-frequency Bursts: in most cases, they were predicted as Low-frequency Lines. The identification of Low-frequency Bursts was lower than for Helix glitches due to the absence of Low-frequency Lines overlapping with \acp{GW} in the training set. 

The Wandering Line glitches, shown in Figure~\ref{fig:other-spectrograms}e, closely resemble the Scratchy type. This morphological similarity may explain why the \ac{LM} classifier identified more than 87\% of these samples as glitches. Finally, we note that \texttt{GSpyNetTree-O4} performed poorly on high-frequency glitch classes such as Whistles (Figure~\ref{fig:other-spectrograms}f), $1400\;\mathrm{Hz}$ Ripples, and $1080\;\mathrm{Hz}$ lines. Since \texttt{GSpyNetTree-O4} was not trained on noise at such frequencies, it predicted the No Glitch class in more than half of the samples for all three classifiers. One solution to this would be to include a class for high-frequency glitches. Modifying \texttt{GSpyNetTree}'s architecture to predict excess power rather than \acp{GW} and specific glitch classes could also improve the identification of high-frequency noise. This direction has been explored for \texttt{GSpyNetTree-S}~\cite{Chan:2025zab}.

\subsection{Effect of the spectrogram Q-value on \texttt{GSpyNetTree-O4}'s classification performance}\label{sec:q-value-val}

\begin{figure}[htbp]
\centering
\captionsetup[subfigure]{font=small}
    \begin{subfigure}[b]{\textwidth}
         \includegraphics[width=\textwidth]{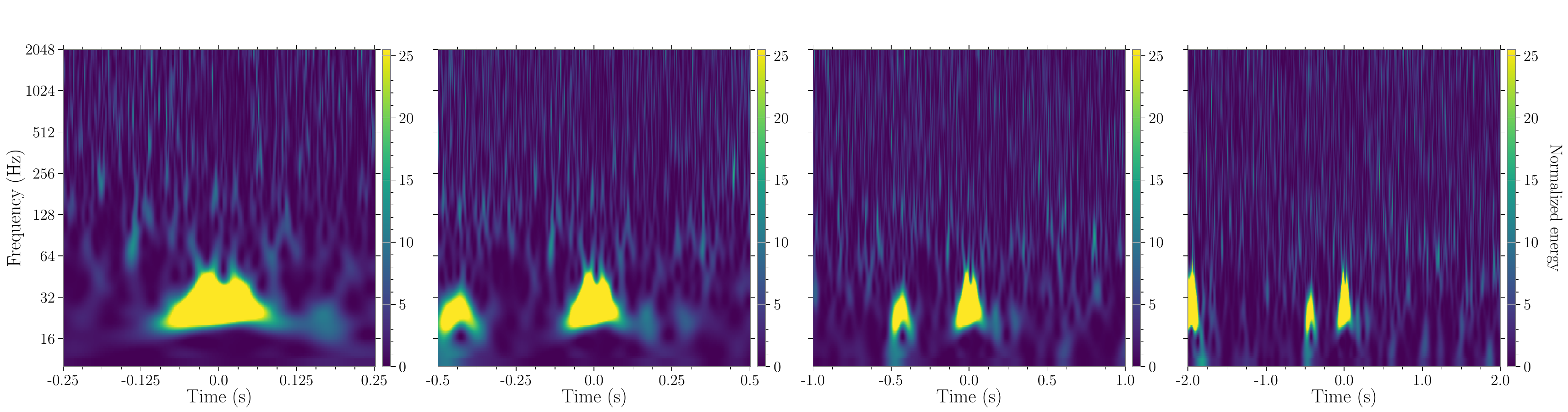}
         \caption[]{} 
         \label{fig:q-value5}
     \end{subfigure}
     \hfill
     \begin{subfigure}[b]{\textwidth}
         \includegraphics[width=\textwidth]{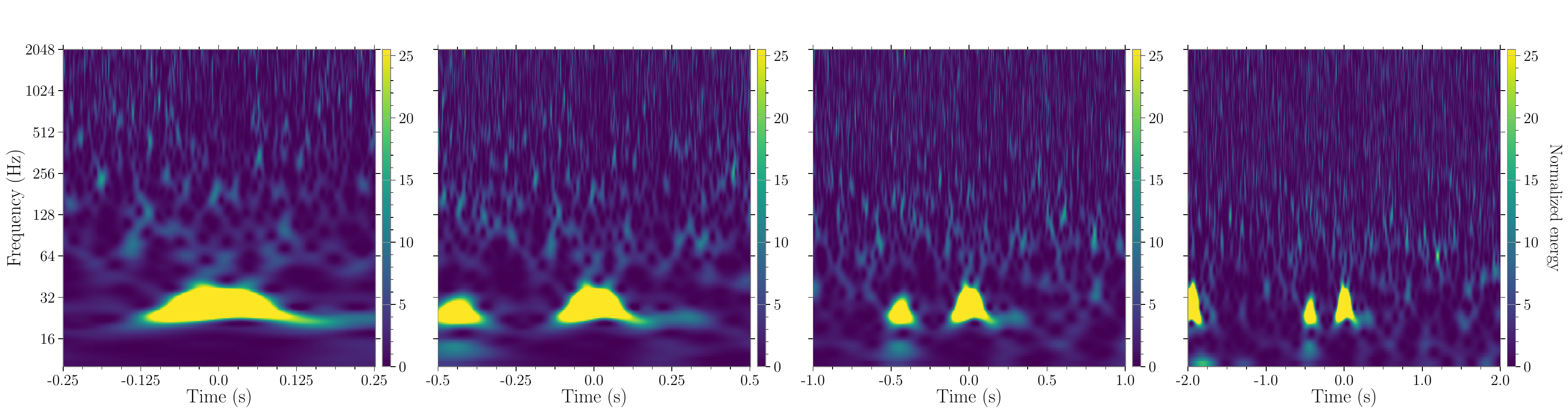}
         \caption[]{} 
         \label{fig:q-value10}
     \end{subfigure}
     \hfill
     \begin{subfigure}[b]{\textwidth}
         \includegraphics[width=\textwidth]{Figures/fast-scattering.pdf}
         \caption[]{} 
         \label{fig:q-value20}
     \end{subfigure}
     \hfill
     \begin{subfigure}[b]{\textwidth}
         \includegraphics[width=\textwidth]{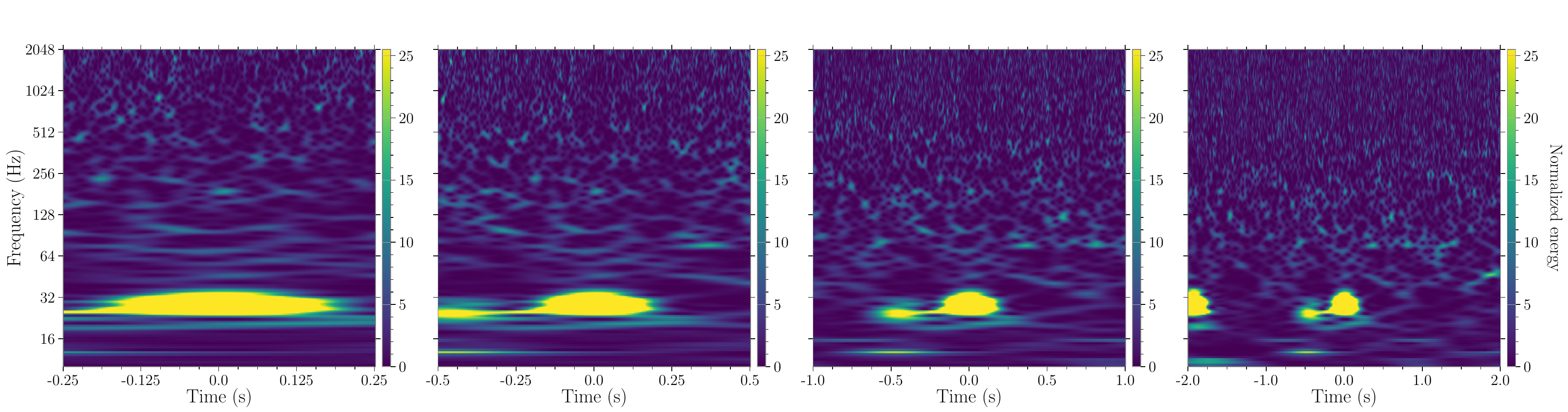}
         \caption[]{}
         \label{fig:q-value30}
     \end{subfigure}
     \hfill
        \caption{Fast Scattering glitch from LIGO Livingston at $Q$-values of (a) $Q=5$, (b) $Q=10$, (c) $Q=20$, (d) $Q=30$. Note that higher (lower) $Q$-values yield better frequency (time) resolution.}
        \label{fig:different-q-values}
\end{figure}

We studied the effect of the spectrogram $Q$-value on glitch and \ac{GW} classification performance for \texttt{GSpyNetTree-O4}. The $Q$-value determines the time-frequency resolution of the spectrogram: higher (lower) $Q$ yields finer frequency (time) resolution~\cite{chatterji}. Figure~\ref{fig:different-q-values} shows the effect of the $Q$-value on the spectrogram representation of a Fast Scattering glitch for $Q \in [5, 10, 20, 30]$. Clearly, the time-frequency morphology of the glitch varies considerably with the $Q$-value (see Section 2 in~\textcite{Ferreira:2024gzh} for the impact of the $Q$-value in unsupervised machine learning methods).

To determine the optimal $Q$-value for each classifier, we regenerated the training set described in Section~\ref{sec:methods} with $Q$-values of 5, 10, 20, and 30, and retrained the \ac{LM} and \ac{HM} classifiers. We excluded Low-frequency Lines from the analysis, as Scattering glitches are similar to Low-frequency Line glitches in duration and frequency content. We therefore expected the impact of the $Q$-value on the time–frequency representation to be comparable. Additionally, we did not perform this analysis for the \ac{EHM} classifier, as its glitch classes are a subset of those included in the \ac{HM} classifier. 

\begin{figure}[!htb]
\centering
\captionsetup[subfigure]{font=small}
     \begin{subfigure}[b]{0.49\textwidth}
         \includegraphics[width=\textwidth]{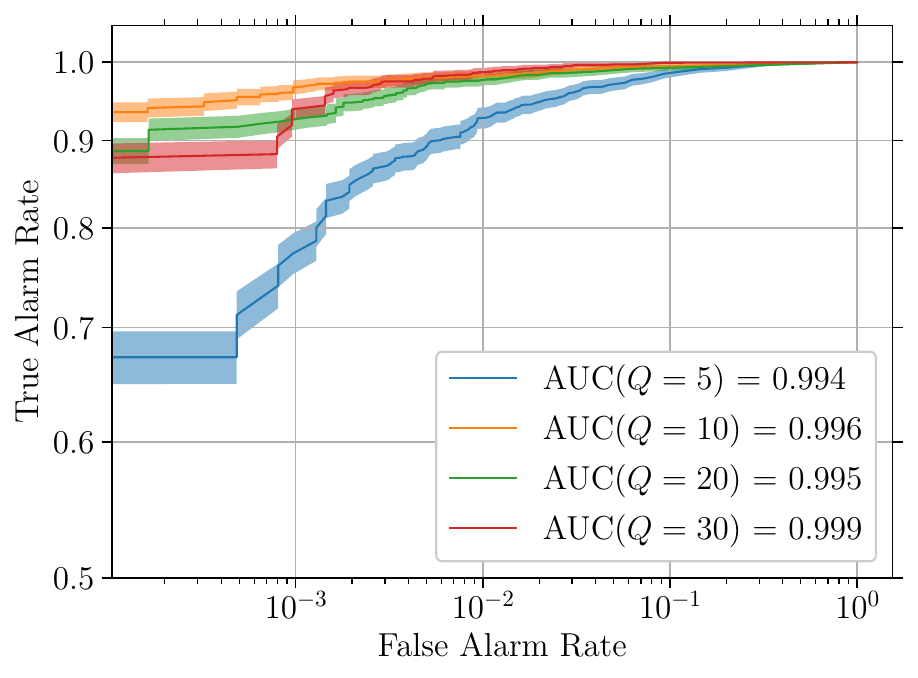}
         \caption[]{}
         \label{fig:roc-lm}
     \end{subfigure}
     \begin{subfigure}[b]{0.49\textwidth}
         \includegraphics[width=\textwidth]{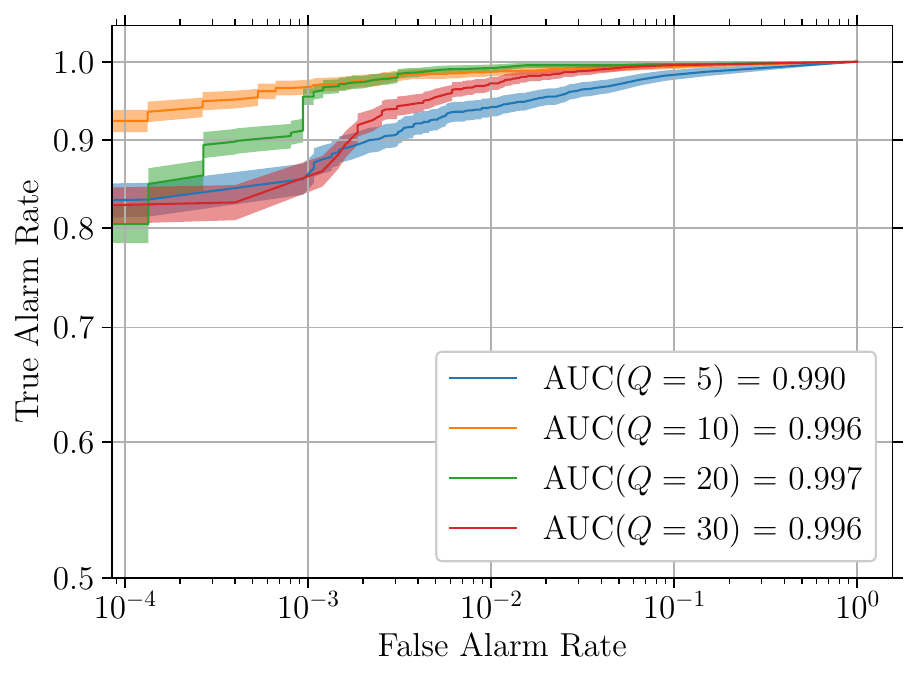}
         \caption[]{}
         \label{fig:roc-hm}
     \end{subfigure}

        \caption{\ac{ROC} curves for the Blip class in the (a) \ac{LM} and (b) \ac{HM} classifiers. The shaded regions represent pointwise $\pm1\sigma$ binomial uncertainties on the \acf{TAR} at each classification threshold. The \ac{AUC} calculated for all $Q$-values is at least $0.99$ in all cases, and the values differ only at the third decimal place. This shows the robustness of \texttt{GSpyNetTree-O4} to different time-frequency resolutions.   } 
        \label{fig:ROC-curves}
\end{figure}

To assess \texttt{GSpyNetTree-O4}'s performance in classifying samples for each $Q$-value, we built \ac{ROC} curves for \acp{GW} and glitches in both the \ac{LM} and \ac{HM} classifiers. \ac{ROC} curves illustrate the trade-off between \acp{TAR} and \acp{FAR} across various classification thresholds, which can be quantified by calculating the \ac{AUC}. Larger AUC indicates better class discrimination, such that a perfect classifier would have an \ac{AUC} of 1. Figure~\ref{fig:ROC-curves} shows an example of the \ac{ROC} curves for the Blip glitch in the \ac{LM} (left) and \ac{HM} (right) classifiers, along with the calculated AUC, for the considered $Q$-values. Note that the \acp{AUC} are very close to 1 for all $Q$-values.

Table~\ref{tab:q-values-both} shows the performance, as measured by \ac{AUC}, for all glitches and \acp{GW} at $Q \in [5,10,20,30]$ in the \ac{LM} (left) and \ac{HM} (right) classifiers, respectively. Across glitch classes, \ac{AUC} remains close to 1 for all tested $Q$-values, whereas the GW-class \ac{AUC} is approximately 0.96 in both classifiers. This slightly lower \ac{GW} performance is expected due to the similarity between low \ac{SNR} signals and No Glitch. Therefore, to select the optimal $Q$-value for each classifier, we considered both the \ac{GW} and mean \ac{AUC} across all classes listed in Table~\ref{tab:q-values-both}.

For the \ac{LM} classifier, the highest \ac{GW} and average \acp{AUC} were achieved at $Q=20$, which was the value used to generate \texttt{GSpyNetTree-O4}’s data sets, described in Section~\ref{sec:methods}. Therefore, this $Q$-value was the optimal for the \ac{LM} classifier. In the \ac{HM} classifier, the average \ac{AUC} was the same at $Q=20$ and $Q=30$. However, the GW \ac{AUC} was $0.3\%$ higher at $Q=30$, which made it the optimal $Q$-value. While the improvement in \ac{AUC} at $Q=30$ in the HM classifier should be considered for future observing runs, we kept $Q=20$ as the default in O4 since it offered a good balance between time-frequency resolution across the three \texttt{GSpyNetTree} classifiers.

\begin{table*}[ht]
\centering
\scriptsize
\begingroup
\setlength{\tabcolsep}{2pt}
\renewcommand{\arraystretch}{1.15}

\begin{minipage}[t]{0.42\textwidth}
\centering
\begin{tabular}{|c|cccc|}
\hline
\multirow{2}{*}{\textbf{\begin{tabular}[c]{@{}c@{}}GSpyNetTree\\ label\end{tabular}}} &
  \multicolumn{4}{c|}{\textbf{AUC} (LM classifier)} \\ \cline{2-5}
 &
  \multicolumn{1}{c|}{\textbf{$Q=5$}} &
  \multicolumn{1}{c|}{\textbf{$Q=10$}} &
  \multicolumn{1}{c|}{\textbf{$Q=20$}} &
  \textbf{$Q=30$} \\ \hline
Blip & \multicolumn{1}{c|}{0.994} & \multicolumn{1}{c|}{0.996} & \multicolumn{1}{c|}{0.995} & 0.999 \\ \cline{1-5}
\begin{tabular}[c]{@{}c@{}}Low-frequency \\ Blip\end{tabular} & \multicolumn{1}{c|}{0.996} & \multicolumn{1}{c|}{0.999} & \multicolumn{1}{c|}{0.997} & 0.992 \\ \cline{1-5}
Fast Scattering & \multicolumn{1}{c|}{0.984} & \multicolumn{1}{c|}{0.991} & \multicolumn{1}{c|}{0.995} & 0.994 \\ \cline{1-5}
Light Scattering & \multicolumn{1}{c|}{0.991} & \multicolumn{1}{c|}{0.993} & \multicolumn{1}{c|}{0.996} & 0.998 \\ \cline{1-5}
Koi Fish & \multicolumn{1}{c|}{1.000} & \multicolumn{1}{c|}{0.999} & \multicolumn{1}{c|}{1.000} & 1.000 \\ \cline{1-5}
Scratchy & \multicolumn{1}{c|}{0.988} & \multicolumn{1}{c|}{1.000} & \multicolumn{1}{c|}{0.996} & 1.000 \\ \cline{1-5}
No Glitch & \multicolumn{1}{c|}{0.961} & \multicolumn{1}{c|}{0.962} & \multicolumn{1}{c|}{0.982} & 0.951 \\ \cline{1-5}
GW ($3-50\;M_\odot$) & \multicolumn{1}{c|}{0.944} & \multicolumn{1}{c|}{0.936} & \multicolumn{1}{c|}{\cellcolor[HTML]{FFFE65}0.949} & 0.947 \\ \cline{1-5}
\textbf{Average} & \multicolumn{1}{c|}{0.982} & \multicolumn{1}{c|}{0.985} & \multicolumn{1}{c|}{\cellcolor[HTML]{FFFE65}0.989} & 0.985 \\ \cline{1-5}
\end{tabular}
\end{minipage}
\hspace{0.035\textwidth}
\begin{minipage}[t]{0.42\textwidth}
\centering
\begin{tabular}{|c|cccc|}
\hline
\multirow{2}{*}{\textbf{\begin{tabular}[c]{@{}c@{}}GSpyNetTree\\ label\end{tabular}}} &
  \multicolumn{4}{c|}{\textbf{AUC} (HM classifier)} \\ \cline{2-5}
 &
  \multicolumn{1}{c|}{\textbf{$Q=5$}} &
  \multicolumn{1}{c|}{\textbf{$Q=10$}} &
  \multicolumn{1}{c|}{\textbf{$Q=20$}} &
  \textbf{$Q=30$} \\ \hline
Blip & \multicolumn{1}{c|}{0.990} & \multicolumn{1}{c|}{0.996} & \multicolumn{1}{c|}{0.997} & 0.996 \\ \cline{1-5}
\begin{tabular}[c]{@{}c@{}}Low-frequency \\ Blip\end{tabular} & \multicolumn{1}{c|}{0.991} & \multicolumn{1}{c|}{0.992} & \multicolumn{1}{c|}{0.996} & 0.989 \\ \cline{1-5}
Fast Scattering & \multicolumn{1}{c|}{0.994} & \multicolumn{1}{c|}{0.991} & \multicolumn{1}{c|}{0.996} & 0.995 \\ \cline{1-5}
Light Scattering & \multicolumn{1}{c|}{0.996} & \multicolumn{1}{c|}{0.998} & \multicolumn{1}{c|}{0.995} & 0.999 \\ \cline{1-5}
Koi Fish & \multicolumn{1}{c|}{0.993} & \multicolumn{1}{c|}{0.990} & \multicolumn{1}{c|}{0.997} & 0.998 \\ \cline{1-5}
Tomte & \multicolumn{1}{c|}{0.997} & \multicolumn{1}{c|}{0.996} & \multicolumn{1}{c|}{0.999} & 0.998 \\ \cline{1-5}
No Glitch & \multicolumn{1}{c|}{0.992} & \multicolumn{1}{c|}{0.972} & \multicolumn{1}{c|}{0.987} & 0.992 \\ \cline{1-5}
GW ($50-250\;M_\odot$) & \multicolumn{1}{c|}{0.966} & \multicolumn{1}{c|}{0.963} & \multicolumn{1}{c|}{0.966} & \cellcolor[HTML]{FFFE65}0.969 \\ \cline{1-5}
\textbf{Average} & \multicolumn{1}{c|}{0.990} & \multicolumn{1}{c|}{0.987} & \multicolumn{1}{c|}{\cellcolor[HTML]{FCFF2F}0.992} & \cellcolor[HTML]{FCFF2F}0.992 \\ \cline{1-5}
\end{tabular}
\end{minipage}

\endgroup

\vspace{0.5em}
\caption{Performance of \texttt{GSpyNetTree-O4}'s \ac{LM} (left) and \ac{HM} (right) classifiers, trained on spectrograms generated using different $Q$-values, as measured by \acf{AUC}. The shaded cells show the highest \ac{GW} \ac{AUC} and average \ac{AUC} for each classifier.}
\label{tab:q-values-both}
\end{table*}

\subsection{Testing \texttt{GSpyNetTree-O4} on a set of Virgo glitches from O4}\label{sec:virgo-glitches-val}

Finally, we evaluated \texttt{GSpyNetTree-O4}'s performance on a set of 200 \ac{O4} Virgo glitches, selected from representative samples of the most prominent Virgo glitch classes in O4: ``25 minute'' glitches and Scattered Light. The samples were manually assigned to the closest \texttt{GSpyNetTree-O4} morphology classes. Although this test was not intended to provide a comprehensive assessment of \texttt{GSpyNetTree-O4}'s applicability to Virgo data, it served as a preliminary study of its transferability. As detailed in Section~\ref{sec:methods}, \texttt{GSpyNetTree-O4}'s design considerations were specifically targeted towards LIGO: most of our glitches and time series data came from \ac{LHO} and \ac{LLO}. In addition, we applied LIGO's \ac{O4} method of subtracting power artifacts and harmonics to the samples in our data set. However, our data set also contained Virgo data from \ac{O3}. Therefore, we wanted to test whether \texttt{GSpyNetTree-O4} could generalize to the most frequent Virgo glitches in \ac{O4}, which have morphologies different from those of LIGO glitches.

\begin{figure}
\centering
\includegraphics[width=0.8\linewidth]{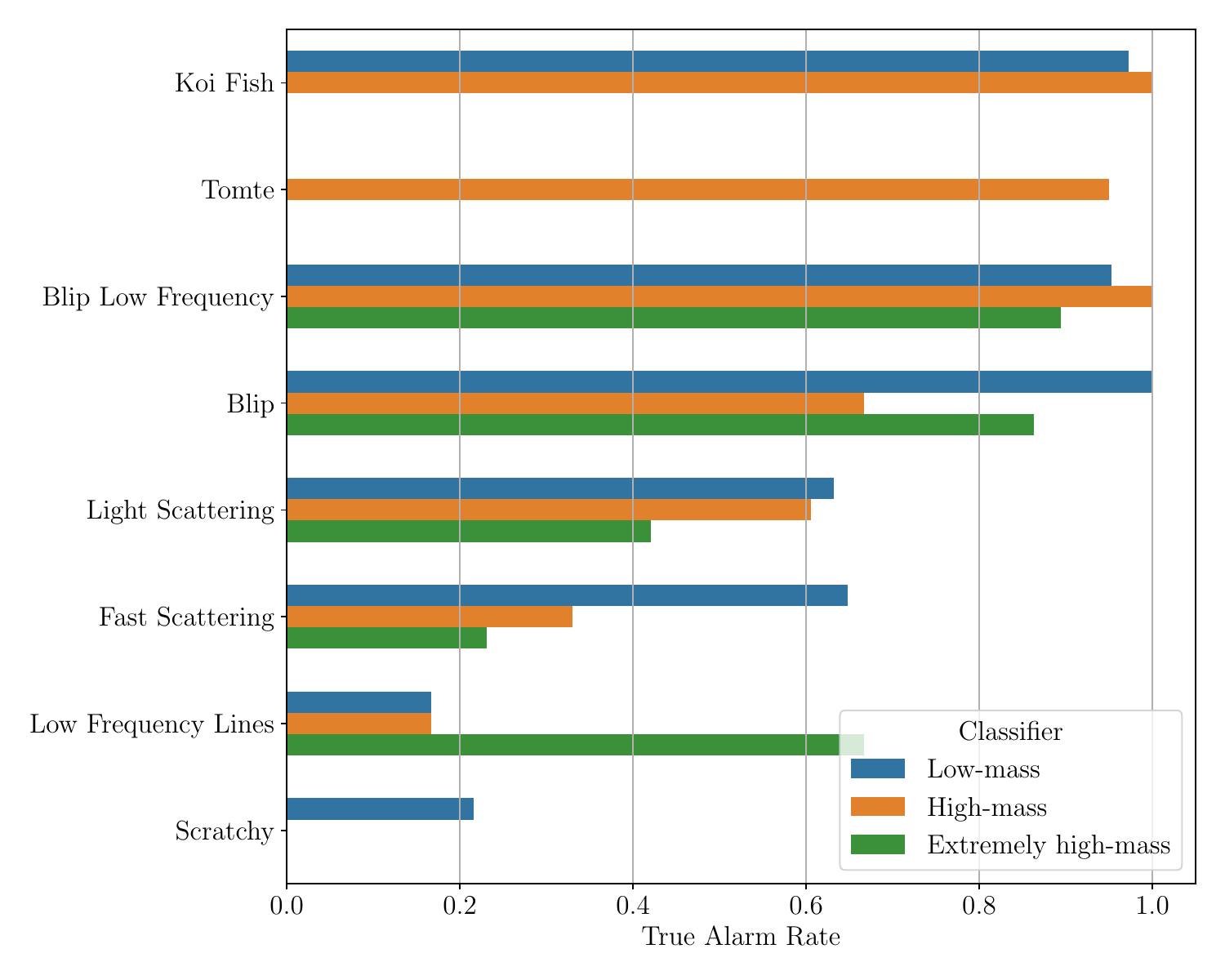}
\caption{Fraction of samples from the Virgo glitch set classified as being/containing a glitch in the \ac{LM} (blue), \ac{HM} (orange), and \ac{EHM} (green) classifiers. More than $90\%$ of Koi Fish, Tomte, and Low-frequency Blip glitches are identified as glitches, while fewer than $60\%$ of Light Scattering, Fast Scattering, Low-frequency Lines, and Scratchy glitches are identified as glitches.
} 
\label{fig:virgo-glitches}
\end{figure}

Figure~\ref{fig:virgo-glitches} shows the fraction of Virgo test samples classified as containing a glitch, regardless of whether a \ac{GW} label was also predicted or whether the predicted glitch type was correctly identified by the classifiers. \texttt{GSpyNetTree-O4} correctly detected a glitch in $\gtrsim 90\%$ of Koi Fish, Tomte, and Low-frequency Blip samples. For the Blip class, the \ac{LM} classifier was able to identify all tested samples. However, the \ac{HM} and \ac{EHM} classifiers missed 33\% and 14\% of the tested samples. For all other classes (Light Scattering, Fast Scattering, Low-frequency Lines, and Scratchy), the glitch identification rate was $\lesssim 60\%$. 

Overall, this performance was substantially below that obtained by \texttt{GSpyNetTree-O4} on its test set, for which more than 95\% of glitches were correctly detected (see Figure~\ref{fig:binary-matrices}). Across the Virgo test set, more than $34\%$, $42\%$, and $51\%$ of the samples analyzed by the \ac{LM}, \ac{HM}, and \ac{EHM} classifiers, respectively, were incorrectly classified as containing no glitch. These false-negative rates are considerably higher than those shown in Figure~\ref{fig:binary-matrices} (2.1\%, 2.3\%, and 4.6\% for each classifier, respectively).

These results suggest that a robust application to Virgo data would require a larger, more representative Virgo-specific training set. A representative Virgo training set may include samples of low-frequency noise, such as the one shown in Figure~\ref{fig:specs}b. In Section~\ref{sec:noiseRobustness}, we found that such low-frequency noise was common in Virgo during \ac{O3}, as first reported in~\textcite{Acernese_2023}. Future developments could consider building larger Virgo glitch data sets using Virgo-focused pipelines, such as \texttt{VIGILant}~\cite{Fernandes:2026pis}.

\section{Conclusions}\label{sec:conclusions}

In this paper, we described \texttt{GSpyNetTree-O4}, a machine learning algorithm for \ac{LVK} event validation that was deployed in production as a \ac{DQR} tool during \ac{O4}. Building upon \texttt{GSpyNetTree}, \texttt{GSpyNetTree-O4} consisted of an ensemble of three \acp{CNN}, referred to as the \acf{LM}, \acf{HM}, and \acf{EHM} classifiers, respectively. These classifiers, based on the InceptionV3 architecture~\cite{inceptionv3}, were designed to target \ac{GW} candidates with total mass in different ranges as well as glitches morphologically similar to \acp{GW} in each corresponding mass range.

To develop \texttt{GSpyNetTree-O4}, we implemented a series of modifications and improvements relative to \texttt{GSpyNetTree}, for both data generation and the classifier architecture. First, we expanded the set of background time series segments from \ac{LHO}, \ac{LLO} and Virgo by verifying the Gaussianity of the segments. This allowed us to not only add simulated \ac{GW} signals in more representative background noise, but also to increase the number of glitches in our data set. Moreover, the most significant upgrade for data generation included the construction of a data set where glitches occurred in the proximity of \ac{GW} signals. This data set was designed to mimic the increasingly likely scenarios where \ac{GW} signals overlap with glitches as detector sensitivity increases. In addition, we applied calibration at 60\,Hz to remove line artifacts, doubling the number of \ac{LHO} and \ac{LLO} samples in our data set. These improvements and modifications allowed us to construct a data set that was more representative of the data expected in \ac{O4}. Finally, to enable simultaneous identification of \acp{GW} and glitches in a single input, we implemented a multi-label architecture for the \ac{LM}, \ac{HM} and \ac{EHM} classifiers of \texttt{GSpyNetTree-O4}, which was one of the most significant upgrades introduced in this work.

On the test sets, \texttt{GSpyNetTree-O4} achieved glitch true-alarm rates $>95$\%. Among samples with no true \ac{DQ} issue, including \ac{GW}-only and No Glitch samples, \texttt{GSpyNetTree-O4} correctly reported no \ac{DQ} issue in 97.1\%, 96.6\%, and 96.0\% of cases for the \ac{LM}, \ac{HM}, and \ac{EHM} classifiers, respectively. In the case of \acp{GW}, particularly in the \ac{LM} regime, low-SNR signals were often misclassified as No Glitch. Likewise, when a low-SNR \ac{GW} signal occurred near a glitch, \texttt{GSpyNetTree-O4} typically detected only the glitch. Although this was sufficient for \texttt{GSpyNetTree-O4}'s use case (i.e., identifying glitches responsible for or in the proximity of candidate events), in future work, we plan to develop improved classification techniques for fainter \ac{GW} signals.

We further assessed \texttt{GSpyNetTree-O4}'s ability to detect glitch types that were not included in the training set. We found that it generalized well to glitches that were morphologically similar to those in its training set, correctly identifying more than 95\% of them as glitches. We also found similar performance when testing \texttt{GSpyNetTree-O4} on glitches characterized by repeated excess power with morphologies resembling those in the training set (e.g., Repeating Blips). However, we found poor performance on high-frequency glitches, primarily because \texttt{GSpyNetTree-O4}'s training set did not include these types of glitches.

We also evaluated \texttt{GSpyNetTree-O4}'s performance on a small set of Virgo glitches from \ac{O4} as a preliminary study of its transferability to Virgo data in future observing runs. Because \texttt{GSpyNetTree-O4} was developed primarily for LIGO data, most of the training samples came from the LIGO detectors. Additionally, the tested Virgo glitches were morphologically different from glitches in the training set. As a result, \texttt{GSpyNetTree-O4} did not generalize well to the tested Virgo glitches. However, the development of a training set with sufficient and representative samples from Virgo is expected to improve the performance of \texttt{GSpyNetTree-O4} for Virgo.

Moreover, we investigated the impact of different $Q$-values used for generating the training set for the \ac{LM} and \ac{HM} classifiers. We found that the $Q$-value employed in this work (i.e., 20) was the optimal for the \ac{LM} classifier among the tested values, while the optimal value for the \ac{HM} classifier was 30. Overall, the effect of the $Q$-value was small.

During O4, \texttt{GSpyNetTree-O4}  was successfully deployed as a \ac{DQR} tool. It demonstrated promising classification performance and increased the degree of automation available to \ac{GW} event validation workflows.

\ack{
We thank Derek Davis and Heather Fong for useful discussions.
The authors are grateful to Francesco DiRenzo for identifying the O4 Virgo glitches used in this paper and for serving as our internal LVK reviewer. 
SA-L. is supported by the Thomas A. Frank Fellowship Fund, the Santodomingo Fellowship Fund, and the ACM SIGHPC Data Science Fellowship. 
MC and JM are supported by an NSERC Discovery Grant, an NSERC Alliance award, CIFAR, and the McDonald Institute. JM is supported by a Tier II CRC. 
DR was supported by a Mitacs Globalink Research Internship award. 
AA was supported by an Erich Vogt First Year Summer Research Experience (FYSRE) award.
JD, AL were supported by NSERC Undergraduate Summer Research Awards (USRAs). 
The authors are grateful for computational resources provided by the LIGO Laboratory supported by National Science Foundation Grants PHY-0757058 and PHY-0823459.
This material is based upon work supported by NSF's LIGO Laboratory which is a major facility fully funded by the National Science Foundation. }

\printbibliography 
\end{document}